\documentclass[iop]{emulateapj}

\usepackage{longtable}
\usepackage{hyperref}
\usepackage{textcomp}
\usepackage{tabularx}
\usepackage{lipsum}
\usepackage{wrapfig}
\usepackage{float}
\usepackage{dblfloatfix}
\usepackage{here}
\usepackage{amsmath}
\usepackage{rotating}
\usepackage{blindtext}

\newcommand{\nkepstars}{99}

\slugcomment{For submission to AAS journals}

\shorttitle{The Binary Companions of Asteroseismic Stars}
\shortauthors{J. S. Schonhut-Stasik et al.}

\begin{document}


\title{Robo-AO \textit{Kepler} Asteroseismic Survey. I. Adaptive optics imaging of 99 asteroseismic \textit{Kepler} dwarfs and subgiants}


\author{Jessica S. Schonhut-Stasik\altaffilmark{1, 2}, Christoph Baranec\altaffilmark{1}, Daniel Huber\altaffilmark{1, 3, 7, 8}, Carl Ziegler\altaffilmark{4}, Dani Atkinson\altaffilmark{1}, Eric Gaidos\altaffilmark{5},  Nicholas M. Law\altaffilmark{4}, Reed Riddle\altaffilmark{6}, Janis Hagelberg\altaffilmark{1}, Nienke van der Marel\altaffilmark{1}, and Klaus W. Hodapp\altaffilmark{1}}
\email{jessicaschonhut@gmail.com}


\altaffiltext{1}{Institute for Astronomy, University of Hawai`i at M\={a}noa, Hilo, HI 96720-2700, USA}
\altaffiltext{2}{University of Hertfordshire, Hatfield, Hertfordshire AL10 9AB, United Kingdom}
\altaffiltext{3}{Sydney Institute for Astronomy (SIfA), School of Physics, University of Sydney, NSW 2006, Australia}
\altaffiltext{4}{Department of Physics and Astronomy, University of North Carolina at Chapel Hill, Chapel Hill, NC 27599-3255, USA}
\altaffiltext{5}{Department of Geology \& Geophysics, University of Hawai`i at M\={a}noa, Honolulu, HI 96822, USA}
\altaffiltext{6}{Division of Physics, Mathematics, and Astronomy, California Institute of Technology, Pasadena, CA 91125, USA}
\altaffiltext{7}{Stellar Astrophysics Centre, Department of Physics and Astronomy, Aarhus University, Ny Munkegade 120, DK-8000 Aarhus C, Denmark}
\altaffiltext{8}{SETI Institute, 189 Bernardo Avenue, Mountain View, CA 94043, USA}


\begin{abstract}

We used the Robo-AO laser adaptive optics system to image \nkepstars{} main sequence and subgiant stars that have \textit{Kepler}-detected asteroseismic signals. Robo-AO allows us to resolve blended secondary sources at separations as close as $\sim$0\farcs15 that may contribute to the measured \textit{Kepler} light curves and affect asteroseismic analysis and interpretation. We report 8 new secondary sources within 4\farcs0 of these \textit{Kepler} asteroseismic stars. We used Subaru and Keck adaptive optics to measure differential infrared photometry for these candidate companion systems. Two of the secondary sources are likely foreground objects while the remaining 6 are background sources; however we cannot exclude the possibility that three of the objects may be physically associated. We measured a range of \textit{i}\textquotesingle{}-band amplitude dilutions for the candidate companion systems from 0.43\% to 15.4\%. We find that the measured amplitude dilutions are insufficient to explain the previously identified excess scatter in the relationship between asteroseismic oscillation amplitude and the frequency of maximum power. 

\end{abstract}
\label{sec:A}




\section{Introduction}
\label{sec:I}

Asteroseismology is the study of the internal structure of stars through the interpretation of their brightness oscillations \citep{Aerts2010}. As early as the 17th century these variations in starlight were identified in stars with large amplitude brightness variations, such as Cepheids and other long-period variable stars \citep{Holwards1642}. Asteroseismic behavior was initially studied only in the fundamental radial mode in which the star maintains spherical symmetry as it expands and contracts. It is now known that many stars pulsate both in radial and non-radial modes, including our own Sun. As well as advancing the understanding of variability, asteroseismology allows the determination of the fundamental properties of these stars, including density, radius, mass and age \citep{Metcalfe2014, Chaplin2014} and has also led to discoveries relating to core properties and rotation rates \citep{Bedding2011, Deheuvels2012}.


It has been claimed that all stars with significant surface convection zones will feature oscillations \citep{Handler2013}. Each observed frequency of oscillation probes an independent measure of an element of the stellar structure, so the more modes for which oscillations are detected, the more information we gather about the star \citep{Christensen-Dalsgaard2014}. Consecutive oscillation modes are often too close in frequency to be distinguished using ground-based observations. Therefore, observations need to take place above the atmosphere in order to resolve these consecutive modes. Many past and present space-based missions have contributed to the field of asteroseismology including: the Wide-field InfraRed Explorer (WIRE) \citep{Hacking1997}, the Microvariability and Oscillations of Stars Telescope (MOST) \citep{Walker2003}, the COnvection, ROtation and planet Transits mission  (CoRoT) \citep{Auvergne2009}, the BRIght Target Explorer (BRITE) \citep{Schwarzenberg-Czerny2010} and the \textit{Kepler} mission \citep{Borucki2010}. \textit{Kepler} alone has revolutionized asteroseismology as a result of its extensive target sample, near continuous monitoring capability and its increased photon collection power. 

Approximately 17,000 stars that demonstrate solar-like oscillations have been observed in short- and long-cadence by the \textit{Kepler} mission \citep{Mosser2012, Hekker2011, Bedding2010, Huber2011, Stello2013}. These data are exploited in a number of ways including the examination of oscillation frequencies in \textit{Kepler} planet host stars \citep{Davies2016} and studying oscillation mode linewidths and mode heights \citep{Appourchaux2014}. The quality of the reduced \textit{Kepler} asteroseismic data is also improving, e.g., by mitigating the impact of the regular gaps in the data \citep{Garcia2014}.

Despite the advantages of \textit{Kepler}, unresolved secondary sources within the \textit{Kepler} photometric apertures can affect our ability to measure asteroseismically determined stellar properties. This occurs when the radiant flux from the secondary star dilutes the frequency amplitudes in the primary light curve. For example, \cite{Campante2014} sets lower limits for the surface gravity of planet candidate host stars, provided no solar-like oscillations are detected. If an unidentified secondary source was masking solar oscillations this would result in an inaccurate lower limit for surface gravity. 

For an asteroseismic star with a physically bound companion, more aspects of the systems can be studied, e.g., \cite{Lai1997}, \cite{Springer2013} and \cite{Polfliet1990} find that tides in close systems can cause the stars to induce or disturb pulsations in one another. Physically associated equal-mass binary stars are likely to follow the same evolutionary path and will display similar oscillation frequencies that may overlap. From the oscillations alone, one may not be able to determine if the target is a binary star or a single hybrid pulsator. 

Adaptive optics (AO) can be used to identify the binary companions and background objects to asteroseismic stars at separations unavailable to spectroscopic or seeing limited observations. \cite{Campante2015} used Robo-AO to measure the amount of dilution of the asteroseismic target Kepler-444 in the visible \textit{Kepler} bandpass by an M dwarf spectroscopic binary system 1\farcs9 away. Further observations by \cite{Dupuy2016} with Keck AO were unable to resolve the individual components of the M dwarf pair and confirmed the absence of additional fainter nearby sources.
 
\protect\footnotetext[1]{In this paper a candidate companion system is used to describe any visible multiple star system, including where the secondary star is an unassociated asterism and a binary system are two physically associated stars. Binary fraction is used to define the number of physically associated stars whereas candidate companion fraction includes all visual candidate companion systems.}

While a handful of asteroseismic stars have been imaged with AO, here we report the first systematic AO survey of asteroseismic stars. We used Robo-AO to observe 99 \textit{Kepler} stars displaying oscillations to determine the existence of any blended binary companions\footnotemark[1] which may be contributing to the stellar light curve. We then used Keck and Subaru AO to obtain differential infrared (IR) photometry in order to determine the spectral types of each candidate companion. A subset of these systems were also observed spectroscopically for more accurate constraints on their spectral types. 

The work is organized as follows: Section \ref{sec:TSAO} describes the target selection and observations. The data reduction and analysis is discussed in Section \ref{sec:DRAA}. Analysis of stellar properties occurs in \ref{sec:SPAPA}. In Section \ref{sec:R} we review the results of the survey, discoveries we have made, follow-up measurements and target confirmation. Section \ref{sec:D} discusses the impact of these results and further analysis. Conclusions and future work are noted in Section \ref{sec:C}. 

\section{Target Selection and Observation}
\label{sec:TSAO}

\subsection{Target Selection}
\label{sec:TS}

To determine if \textit{Kepler} targets demonstrated detectable oscillations, short cadence data of $\sim$2000 solar-like stars were provided to the asteroseismic community \citep{Chaplin2010, Gilliland2010, Huber2011, Verner2011}. Approximately 500 were found to display observable oscillations \citep{Chaplin2010, Verner2011} and the \textit{Kepler} Community Follow-up Observation Program (CFOP) compiled these stars in to a list of \textquotesingle{}standard stars\textquotesingle{}. Each star in the list was assigned a priority of either Platinum or Gold. Platinum stars have full-mission, short-cadence data. Gold stars have lower signal-to-noise asteroseismic detections, e.g., a month of short cadence data. We selected the Platinum sample for our initial AO survey with the intention of observing the full Gold priority sample in later work. Ultimately our target sample contains 97 Platinum and 2 Gold stars due to a reevaluation of the signal-to-noise quality of two targets by CFOP. Seven of the observed platinum standard stars are also \textit{Kepler} Objects of Interest (KOIs), which are stars that show repeating transit signals. The target sample is detailed in Table \ref{long}.

The stars in our sample are solar-like with effective temperatures between 4910-6700K. Solar-like stars are most likely to display stochastic oscillations as opposed to the coherent pulsations seen in hotter stars. These solar-like oscillations are predominantly acoustic waves (p-modes) that propagate via the compression and rarefaction of gas, with the pressure gradient acting as the restoring force. Because oscillations in solar-like main sequence stars usually have periods of $\sim$5 - 13 minutes, multiple periods can be resolved by \textit{Kepler} in short cadence mode. 


\begin{figure}[h]
\includegraphics[height = 7cm]{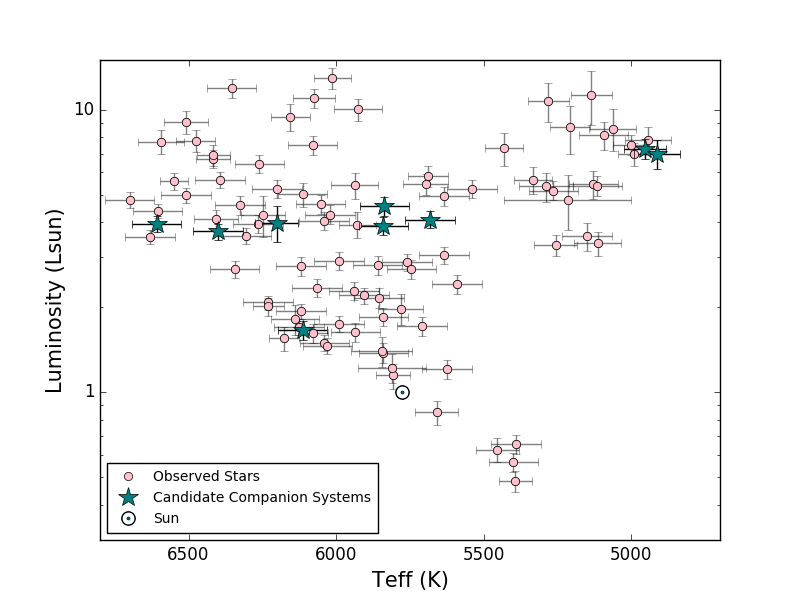}
\caption{H-R diagram of our target sample with green stars showing the 9 candidate companion systems detected. Our sample of stars span a broad area of the lower main sequence/red giant branch. At least two of the candidate companion systems have subgiant primaries which display a high temperature and luminosity. The Sun has been shown for reference.}
\label{fig:HRDia}
\end{figure}

Stars with greater apparent brightness are more easily observed due to increased signal and stars with higher luminosity are more likely to demonstrate observable oscillations because oscillation amplitudes increase proportionally with luminosity \citep{Huber2011, Kjeldsen1995}. The standard stars list contains bright stars relative to the entire \textit{Kepler} Input Catalog (KIC). The stars in our target sample have \textit{i}\textquotesingle{} magnitudes from 7.1 to 11.3. 

Figure \ref{fig:HRDia} shows the sample on a Hertzsprung-Russell (H-R) diagram. We used the CFOP catalog\footnotemark[2] to obtain the targets' effective temperatures and radii, with errors, and with the exception of KIC 8760414. The luminosity errors were determined using the errors in temperature and radius. The majority of target stars fall on the early-to-mid main sequence, or low on the red giant branch.

\protect\footnotetext[2]{https://exofop.ipac.caltech.edu/cfop.php}

\subsection{Observations}
\label{sec:O}

\subsubsection{Robo-AO}
\label{sec:RA}

We used the Robo-AO robotic visible-light laser AO system \citep{Baranec2013, Baranec2014}, mounted on the 1.5m telescope \citep{Cenko2006} at Palomar Observatory, to obtain high angular resolution images of the \nkepstars{} asteroseismic stars comprising the target sample. We used a queue scheduled mode \citep{Riddle2014} to perform all observations contemporaneous with other science programs.  Observations took place between 2014 June 15 and 2014 November 7, across 18 nights, with some objects observed more than once to ensure high quality images. For targets observed multiple times, we indicate the date of the highest quality observation. Table \ref{tab:specs} identifies the survey and system specifications.

The raw images comprise a sequence of full-frame-transfer detector reads from an electron multiplying CCD at a rate of 8.6Hz. We used a total exposure time of 90s that enables the detection of additional sources of up to $\sim$6 magnitudes fainter than the target \citep{Law2014}. We took all observations in the \textit{i}\textquotesingle{} filter to obtain the sharpest possible images, allowing Robo-AO to detect secondary sources within 0\farcs15 for bright targets (m\textless13) in median seeing conditions. 


\begin{table}[h]
\renewcommand{\arraystretch}{1.3}
\begin{longtable}{lc}
\caption{Robo-AO Specifications}
\\
\hline
Sample targets   	& 99 \\
Exposure time & 90s\\
Observation wavelength & $i'$ band \\
FWHM resolution   	& 0$\farcs$15 \\
Field of view & 44\arcsec $\times$ 44\arcsec\\
Pixel scale & 43.1 mas pix$^{-1}$\\
Detector format       	& 1024$^2$ pixels\\
Number of nights & 18  \\
Observation date range &  2014 July 15 -- 2014 November 7 \\
\hline
\label{tab:specs}
\end{longtable}
\end{table}

\subsubsection{Keck Adaptive Optics}
\label{sec:KASFU}

We used the NIRC2 infrared camera behind the Keck II AO system to confirm potential secondary sources and obtain supplementary differential near-infrared photometry. These observations took place on 2016 April 23, 2016 September 12 and 2016 September 13. We operated NIRC2 in its 9.9 mas pixel$^{-1}$ mode which results in a field of view of $\sim$10\farcs0. We used the the Br$\gamma$ filter (central wavelength 2.17$\mu$m) for all observations. We obtained 3-point dithered images for each star with total exposure times from 45s to 180s. 

\subsubsection{Subaru Adaptive Optics}
\label{sec:SAO}

We used the Infrared Camera and Spectrograph (IRCS) behind the Subaru AO system to observe the candidate companions not observed with NIRC2. These observations took place on 2016 June 17. We used IRCS in the 20.57 mas/pixel mode which results in a field of view of 21\farcs06. We took many of these images in poor conditions due to variable weather on the night. We used a K\textquotesingle{} filter (central wavelength 2.12$\mu$m) and obtained multi-point dithered images with total exposure times from 1s to 84s.

\subsubsection{Visible-light spectroscopy}
\label{sec:S}

Integral field, visible-wavelength spectra of 8 candidate systems were obtained with the Super Nova Integral Field Spectrograph (SNIFS) on the UH 2.2m telescope on Maunakea on 2016 April 28 and 2016 May 2. Seeing was typically 0\farcs7-1\farcs1. SNIFS has a $6\arcsec \times 6\arcsec$ field of view ($15\times 15$ pixels) with a 320-970nm spectral range in two channels (B and R) at a resolution of 700-1000 \citep{Aldering2002,Lantz2004}. We used integration times of 1-2 minutes. Flat-field images and Th-Ar spectra were obtained at each pointing to permit accurate calibration of the spectra. 

\section{Data Reduction}
\label{sec:DRAA}

We used the standard Robo-AO data reduction techniques described in \cite{Law2014}. We calibrated and registered individual frames to synthesize a deep exposure image, before running a fully-automated point-spread function (PSF) subtraction and companion detection routine. We used the additional infrared images and photometry to fit a spectral type to each star and calculated the estimated photometric distances for the candidate companion star. We used spectroscopy to determine the spectral types for some stars.

\subsection{Data Reduction and Imaging Pipeline}
\label{sec:DRAIP}

The Robo-AO data reduction pipeline subtracts a dark frame and calibrates the flat-field from each raw frame. We align calibrated frames on the position of the target star then stack them together to create a single reduced image. We manually inspected the reduced images and flagged possible companions within the size of a \textit{Kepler} pixel, $\sim$4\arcsec. We do not report secondary sources at greater separations, as seeing-limited observations can detect these.

\subsection{Imaging Performance Metrics}
\label{sec:IPM}

\begin{figure}[ht]
\includegraphics[height = 7cm]{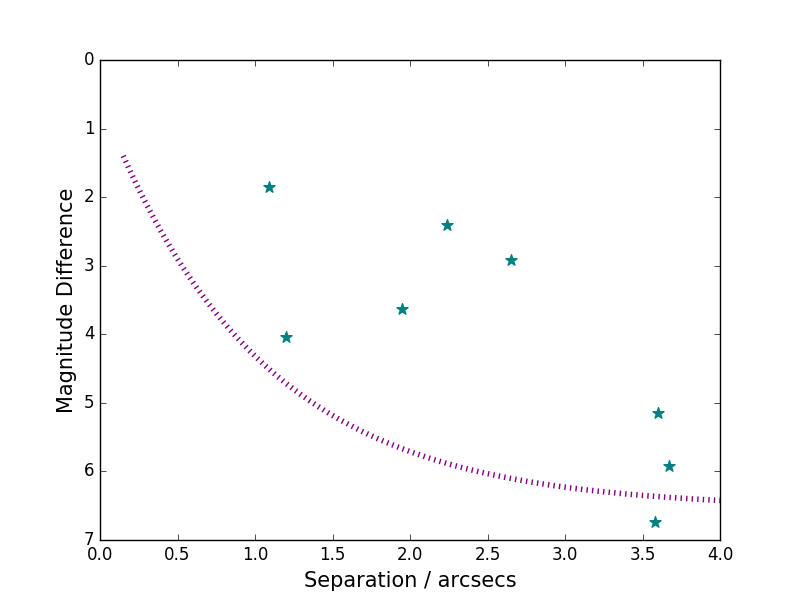}
\caption{Points on this plot show the angular separations and magnitude differences of the detected secondary sources described in Tables \ref{CompDet} and \ref{CompDet2}. The dotted line represents the typical 5$\sigma$ \textit{i}\textquotesingle{}-band contrast curve achieved during the observations with Robo-AO.}
\label{fig:MagDiff}
\end{figure}

Previous analysis of reduced Robo-AO images has shown that the measured width of the core PSF is an excellent indicator of achievable contrast performance \citep{Law2014}. We fit two Moffat functions to the PSF of each observation, one tuned to the PSF core and the other to the uncorrected halo. If the full-width at half-maximum (FWHM) of the core was $\geq$0\farcs14, the image quality and achievable contrast was in the top 30th percentile of all useful Robo-AO observations. If the data obtained for any particular observation did not meet these criteria, the observation was repeated until it did. This ensured homogeneous achievable contrast across the entire survey. The typical $5\sigma$ achievable contrast for Robo-AO images for this survey is shown in Figure \ref{fig:MagDiff}. 


\subsection{PSF Subtraction}
\label{sec:PS}

We used a custom locally optimized PSF subtraction routine based on the Locally Optimized Combination of Images algorithm \citep{lafreniere07} to identify close secondary sources. A reference PSF was created by using a set of twenty observations taken from observations of other stars, imaged in the same filter and closest in observation time to the target observation. The reference PSF was subtracted from the original image, leaving residuals consistent with photon noise.

\subsection{Automated Companion Detection and System Confirmation}
\label{sec:ACDASC}

We used an automated companion detection algorithm developed for our Robo-AO KOI surveys (see \citealt{Ziegler2016}) to detect additional companions not identified in the raw or PSF subtracted data. The significance was found for each candidate companion by sampling and modeling the background noise level as a function of radial distance from the target star. We then slid an aperture of the diffraction-limited FWHM diameter along concentric annuli centered on the target star. Possible astrophysical detections are identified when the enclosed flux of the aperture becomes significantly greater than the local noise. In this sample of brighter stars, bright speckles produce the majority of high-significance detections, which we manually discard. 

\subsection{Reduction of Infrared AO Images}
\label{sec:IRR}

Each dithered image from NIRC2 and IRCS was subject to sky subtraction and flat-field calibration. Each frame was corrected for bad pixels, and stacked to create final images. 
\subsection{Spectroscopy Reduction}
\label{sec:SR}

Extraction and processing of spectra are described in \citet{Aldering2002} and \citet{Mann2015}. Spectra of identified sources were extracted manually and the distance in pixels computed. The position of the primary source in each image \textquotesingle{}slice\textquotesingle{} (specific wavelength) plus the offset was used to place an aperture on each slice allowing us to obtain photometry of the fainter secondary. Aperture photometry was performed on each image slice after subtracting Gaussian fits to the primary. Corrections to the photometry used the atmospheric extinction of \cite{Buton2012} and additional corrections described in \cite{Mann2015}. Spectra were de-reddened and compared (minimizing $\chi^2$) with the G\"ottingen spectral library, which was generated by the PHOENIX model in spherical mode \citep{Husser2013}. Because of low signal in the B channel and lack of photometry to accurately match the two channels, only the R channel was used. Several narrow wavelength ranges that are difficult to model or have strong telluric absorption lines were also excluded \cite[see]{Mann2013}. We adopted the estimated extinction values for each of the targets from the KIC catalog, with the caveat that the actual values could be much larger if the sources are not physical companions.

\section{Analysis of Stellar Properties}
\label{sec:SPAPA}

We measured the relative astrometry and photometry and determined the spectral types for each of the discovered candidate companion systems. We then used this information to determine the distance to each candidate companion star and calculated the possibility of physical association of these systems. We also report the oscillation amplitude dilution due to the radiant flux of the secondary star. These results are summarized in Tables \ref{CompDet} and \ref{CompDet2}.

\subsection{Relative Astrometry}
\label{sec:BCP}

We determined the separation and position angle between the primary and secondary stars, which includes a correction for optical distortion by using the solution produced from Robo-AO measurements of globular clusters detailed in
\cite{Riddle2015}. The PSF subtracted images were used to find the position of the secondary relative to the primary for the companions found at a detection significance of \textless3$\sigma$. 

\subsection{Photometry and Individual Spectral Types}
\label{sec:BCP2}

We used aperture photometry to determine magnitude differences for all the candidate companion systems in both \textit{i}\textquotesingle{} and K\textquotesingle{}. The images are high resolution and with well separated companions, so the brightness of each star can be measured using simple aperture photometry. We corrected for the radiant flux of the primary PSF halo by subtracting an aperture on the opposite side of the candidate companion star.  Errors for the companion photometry were estimated using a varying aperture with the width of the stellar PSF and measuring the standard deviation of the difference.

We compared the spectral types determined through our photometry with the blended system spectral types available on \textit{Kepler} Asteroseismic Science Operations Center (KASOC) online database\footnotemark[4]. We found the spectral types of 6 of the candidate companion systems on KASOC: five F stars and a G star.\protect\footnotetext[4]{http://kasoc.phys.au.dk} The spectral types for the remaining two candidate companion systems were not available so we used apparent magnitudes for J, H, K\textquotesingle{} and i\textquotesingle{} bands from the CFOP catalog and fit them to a catalog of main sequence standards from \cite{Kraus2007} using code developed by \cite{Atkinson2016}. This resulted in two K candidates, KIC 7584900 and KIC 11554100, with effective temperatures of 4960K and 5150K respectively. The primary star dominates the radiant flux in each system and we found that our fit primary spectral types are consistent with the candidate companion system spectral types from KASOC/CFOP. The luminosity classes assigned to these two K subgiants are from asteroseismology.

We used the KIC magnitudes listed on CFOP as the combined apparent magnitude of the blended primary and secondary star. We determined the individual stellar magnitudes for the primary and secondary stars from the combined apparent magnitude and our measured magnitude differences. 

We then used the individual magnitude values to fit the spectral type of each each star. Measured photometry and uncertainties generate Gaussian distributions of i-K and \textit{Kepler}-K. Photometric values are weighted by measured uncertainty, and corrected for reddening using existing $A\_V$ values and standard absorption relations \citep{Cardelli1989}. The \cite{Atkinson2016} model only fits to main sequence stars, does not take into account age or metallicity and does not discriminate between dwarfs and giants. We infer the radius from the spectral fit to the secondary source assuming they are all dwarfs.




\subsection{Amplitude Dilution}
\label{sec:AD}  

Radiant flux from a companion star in an unresolved binary system will contaminate the \textit{Kepler} light curve, reducing the relative amplitude signal, hindering the determination of oscillation frequencies. 
Using the relative photometry between the primary and the secondary stars we determined the amplitude dilution, $A$,

\begin{equation}
A = \frac{f_{2}}{f_{1} + f_{2}},        
\end{equation} in the \textit{i}\textquotesingle{} and K\textquotesingle{} bands, where $f_{1}$ and $f_{2}$ denote the radiant flux of the primary and secondary star respectively. 

\subsection{Estimation of Photometric Distances}
\label{sec:PA}

We used the target distances determined by \cite{Mathur2016}, who revised stellar properties for almost 200,000 \textit{Kepler} stars using isochrone fitting. To determine the corrected distance to the primary, we subtract off the measured radiant flux of the secondary in \textit{i}\textquotesingle{} and recalculate the distance, 

\begin{equation}
d_{corrected} = \sqrt{\frac{1}{1 - A}}d_{blended},       
\end{equation} where $d_{corrected}$ is the corrected distance to the primary star and $d_{blended}$ is the distance reported by \cite{Mathur2016} of the blended system. To calculate distances to the secondary sources, we used the standard distance modulus equation with our measured apparent magnitudes and the absolute magnitudes derived from our fit of spectral type.  

By comparing whether the distances to the primary and secondary stars overlap within measurement error, we are able to determine whether the stars can be physically associated. From the derived distance measurements we calculated the significance of the difference in distances. For values \textgreater3$\sigma$ these systems are likely to be physically unassociated. For a confidence level less than this, we cannot rule out the possibility of physical association.

\section{Discoveries}
\label{sec:R}

\begin{figure}[p]
\centering
\includegraphics[scale = 0.60]{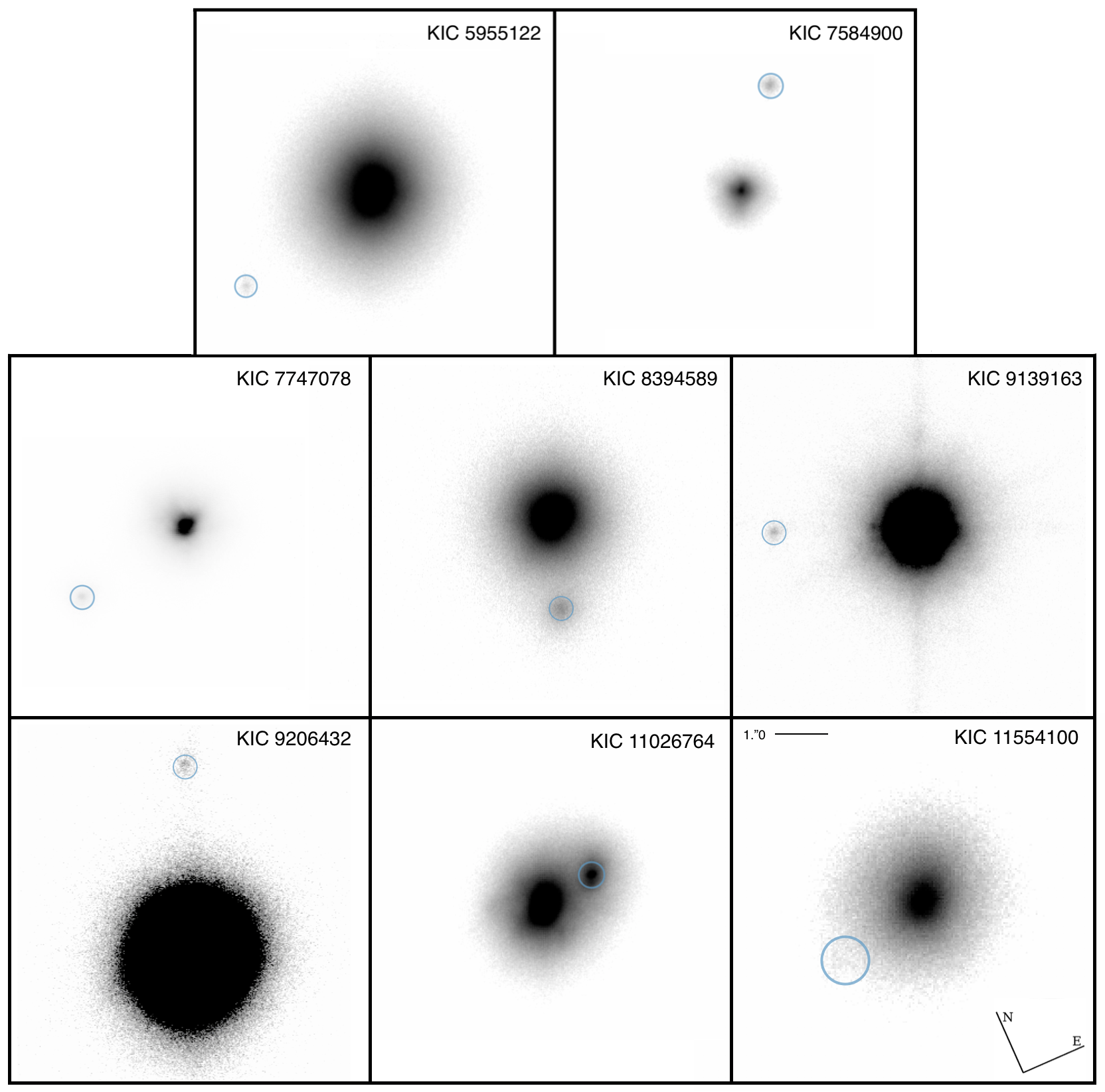}
\caption{Robo-AO \textit{i}\textquotesingle{}-band images of discovered candidate companion systems. The secondary source is outlined in blue circle. Scale and orientation shown at bottom right.}
\label{fig:Discoveries}
\end{figure}

\begin{figure}[p]
\centering
\includegraphics[scale = 0.30]{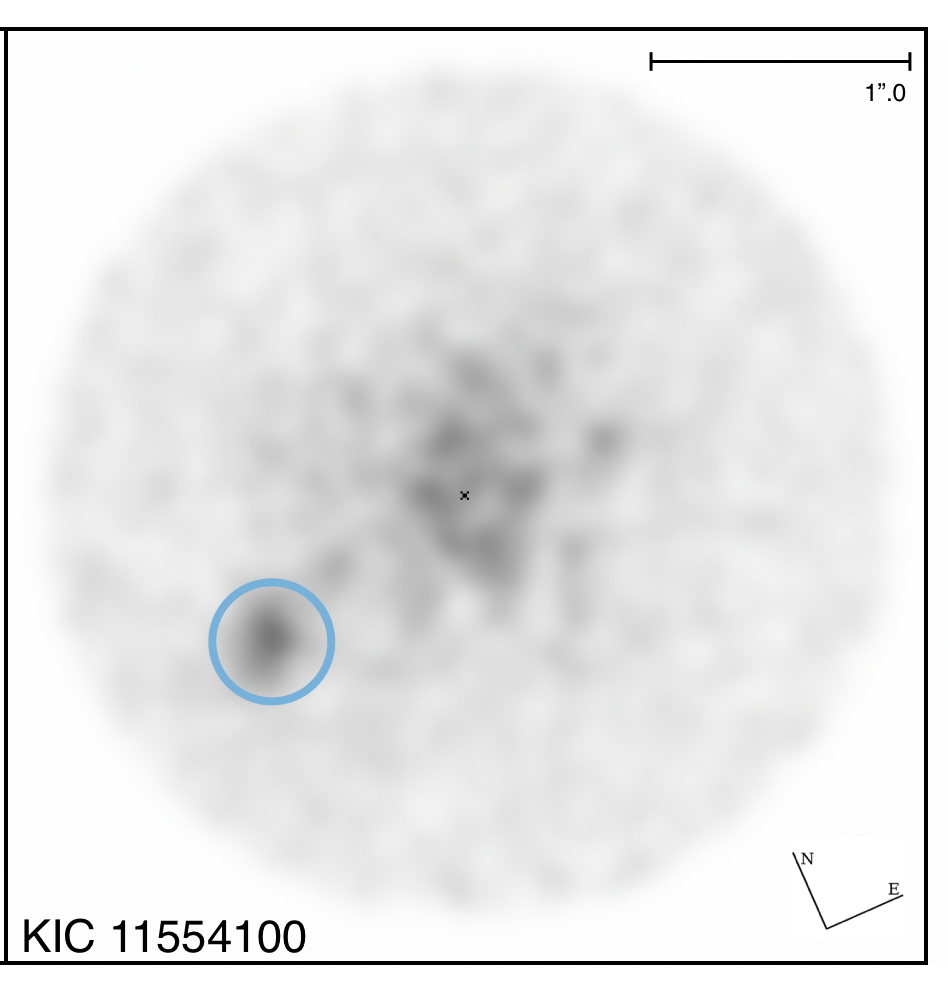}
\caption{PSF subtracted Robo-AO image of KIC 11554100. The image has been convolved with a diffraction-limited Gaussian kernel for clarity. The location of the primary star is indicated with an $\times$.}
\label{fig:psf_fig}
\end{figure}

\begin{figure}[p]
\centering
\includegraphics[scale = 0.45]{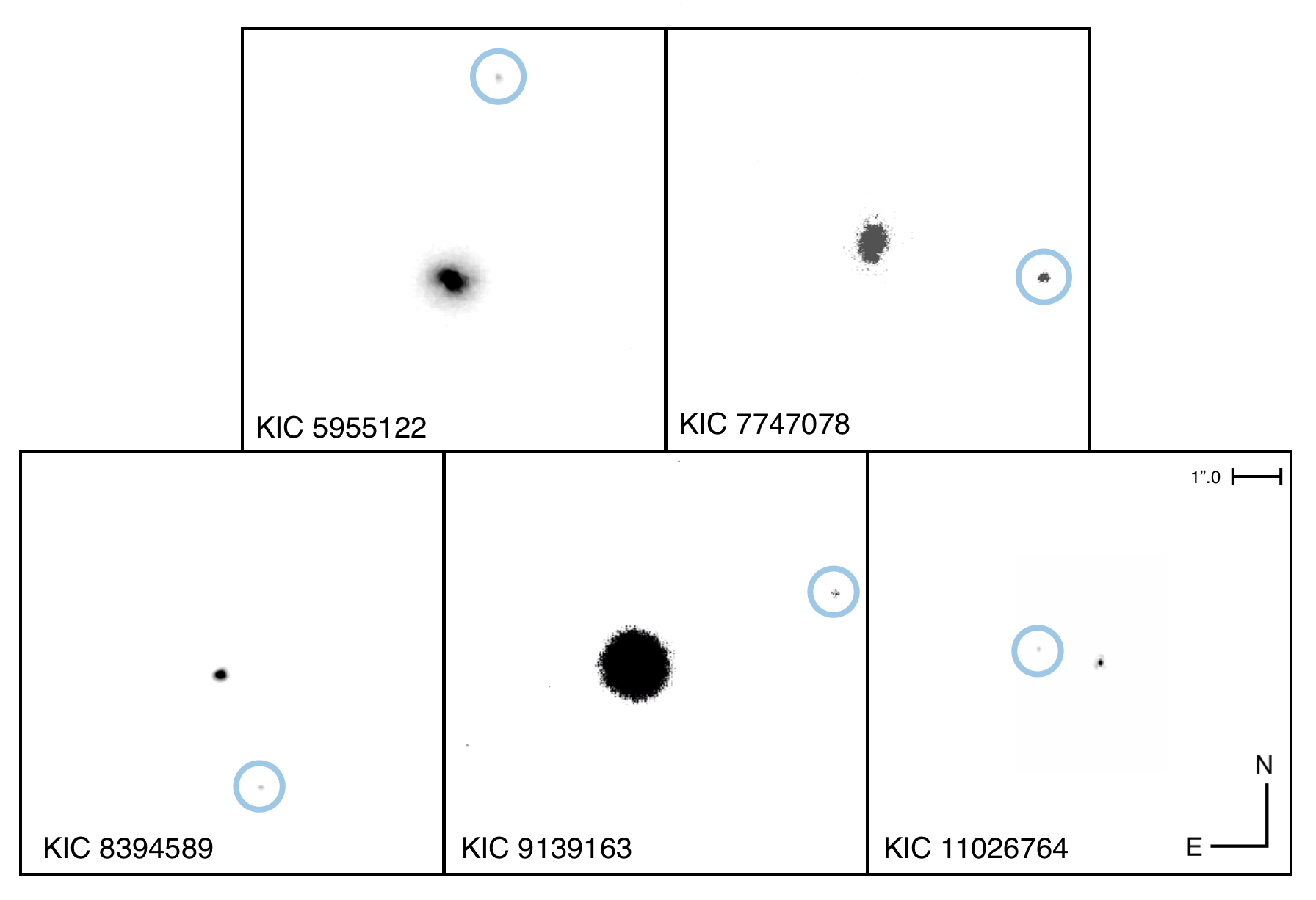}
\caption{IRCS K\textquotesingle{}-band images of stars with detected secondary sources.}
\label{fig:Subi}
\end{figure}

\begin{figure}[p]
\centering
\includegraphics[scale = 0.30]{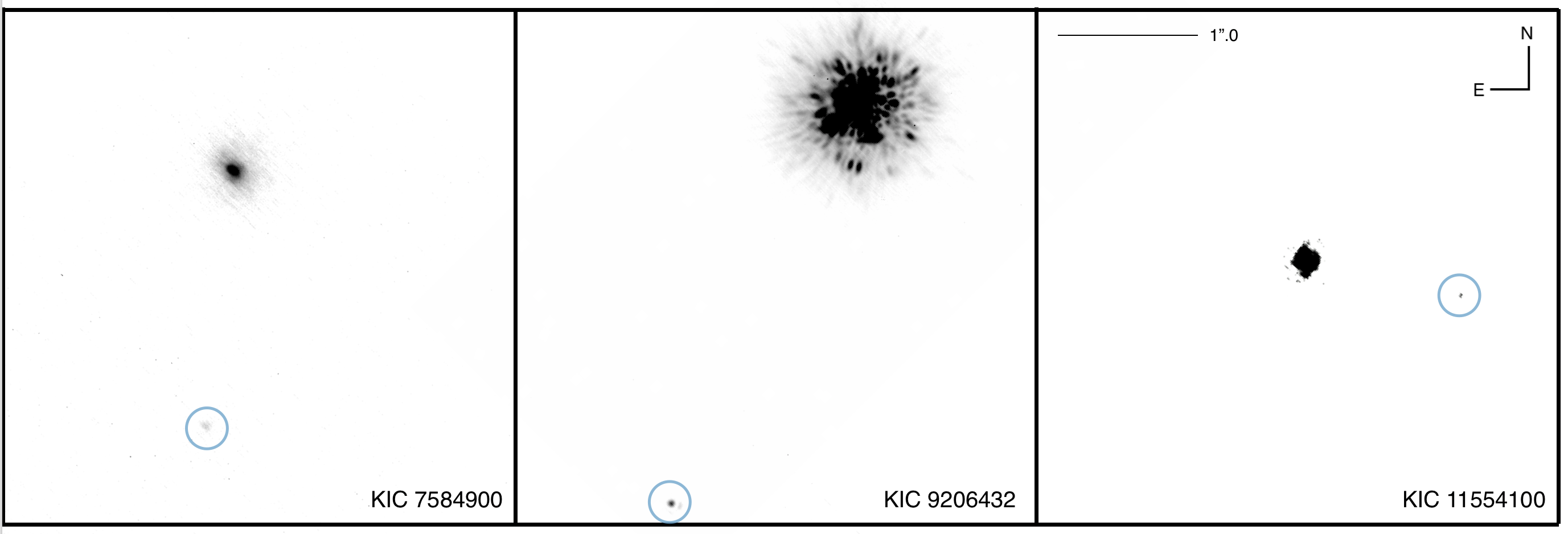}
\caption{NIRC2 Br$\gamma$ images of stars with detected secondary sources.}
\label{fig:Keck}
\end{figure}

We initially found 11 candidate companion sources within 4\farcs0 of the 99 \textit{Kepler} standard stars we observed with Robo-AO. We found 7 in the initial manual search of reduced images and another 4 in the PSF subtracted images. We then used NIRC2 and IRCS to observe all of these candidates (see Table \ref{irobs}) and ruled out three of the candidate secondary sources found from PSF subtracted images. The Robo-AO discovery images are summarized in Figure \ref{fig:Discoveries} with the companion found in the PSF subtracted image, KIC 11554100, shown in Figure \ref{fig:psf_fig}. The IRCS and NIRC2 are shown in Figures \ref{fig:Subi} and \ref{fig:Keck} respectively. Three of the target sample stars, all KOIs, have had secondary sources previously detected that were fainter than our survey sensitivity. These are noted in Table \ref{long}.

\subsection{Spectroscopy}
\label{sec:SRes}

\begin{figure}[p]
\centering
\includegraphics[scale = 0.40]{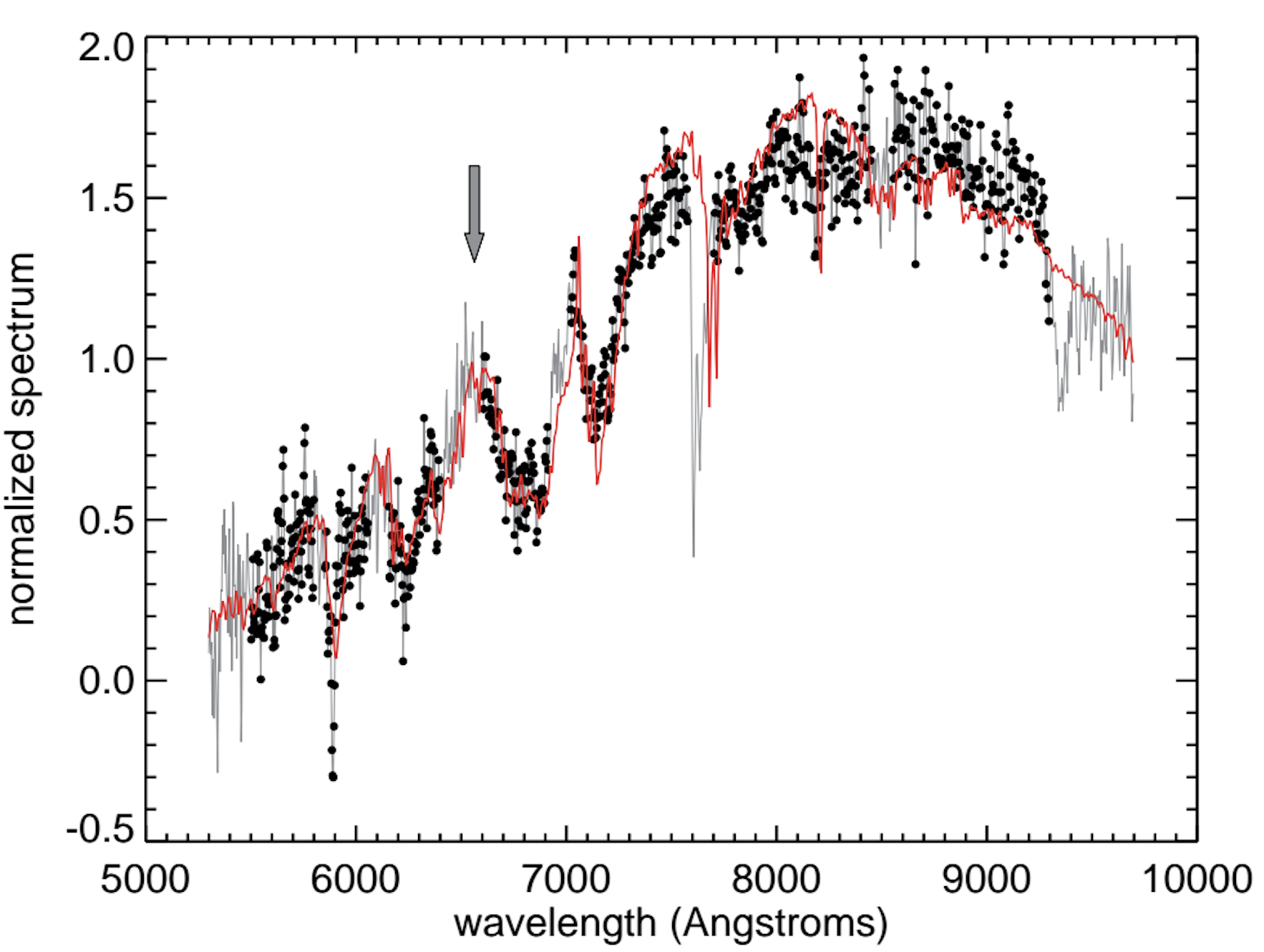}
\caption{SNIFS red-channel spectrum of the putative M dwarf companion of KIC 5955122. The complete spectrum, including telluric-affected regions, is the grey line.  The black points are the points used for fitting, and the red line is the best-fit PHOENIX model spectrum, with Teff=3200K, log g = 5, and [Fe/H] = -0.5. The arrow indicates the location of any possible H$\alpha$ emission.}
\label{fig:Spec}
\end{figure}

We were only able to extract the spectra for 3 of the secondary sources due to decreased achievable contrast ratio with the seeing-limited SNIFS instrument. We find that the secondary to KIC 5955122 has a spectrum consistent with that of a late-type ($T_e \approx 3200$K), metal-poor ([Fe/H] $\approx -0.5$), M dwarf, with an indication of H$\alpha$ in emission (See Figure \ref{fig:Spec}). The secondary sources to KIC 7584900 and 7747078 have spectra that match hotter dwarfs, with $T_e \sim 5900$ and 6500K, respectively, but with large uncertainties.\footnotemark[5] B-band fluxes from both of these stars are significantly attenuated with respect to the best-fit models, suggesting that extinction along the line of sight is much higher and, possibly, the stars are more distant and unrelated to the targets. Values derived from spectroscopy are noted in Table \ref{CompDet2}.

\protect\footnotetext[5]{SNIFS spectra readily resolve the broad molecular features of late K and M dwarfs, but not the atomic lines that are important temperature indicators in the spectra of hotter stars.}

\subsection{Are Secondary Stars Background/Foreground Objects?}
\label{sec:CCS}

We calculated the probability of the 8 candidate companions not being associated using their photometric distances (see Table \ref{CompDet2}). We found six of these candidate companions are likely to be background sources, although we cannot rule out that KICs 9139163 and 11554100 may be potentially physically associated. The remaining two candidate companions are likely foreground objects, however KIC 7584900 may be physically associated. It is possible that these secondary sources could be very distant giants as opposed to nearby dwarfs. 

\section{Discussion}
\label{sec:D}

\subsection{Impact on Stellar Oscillation Amplitudes}
\label{sec:ABIPS}

By studying stellar oscillation amplitudes, we gain insight into the poorly understood physics of surface convection that cause the stochastic excitation and damping of solar-like oscillations \citep[e.g.,][]{houdek99}. Over the last two decades a consensus has emerged that oscillation amplitudes scale as a function of luminosity, temperature and mass with different power law coefficients \citep{Kjeldsen1995,stello11,Huber2011,corsaro13}. Despite the recent advances due to the large sample observed by \textit{Kepler}, the residual scatter of amplitude scaling relations remain significantly larger than the measurement uncertainties. \cite{Huber2011} shows a standard deviation of the oscillation amplitudes that is 2.4 times larger than the median uncertainty. This suggests that the errors do not explain the scatter, which implies an as of yet unidentified additional physical mechanism influencing oscillation amplitudes.

We binned the values of frequency of maximum power from \cite{Huber2011} to widths of 100$\mu$Hz and calculated the ratio of amplitude standard deviation to average measurement error for each bin. All but one bin produced a result greater than 1, with a range of 0.9 - 9.5, implying the scatter of the amplitude measurements exceeds measurement error. The binned values produce a ratio of standard deviation to measurement error of 3.0.

\begin{figure}[]
\includegraphics[height = 7cm]{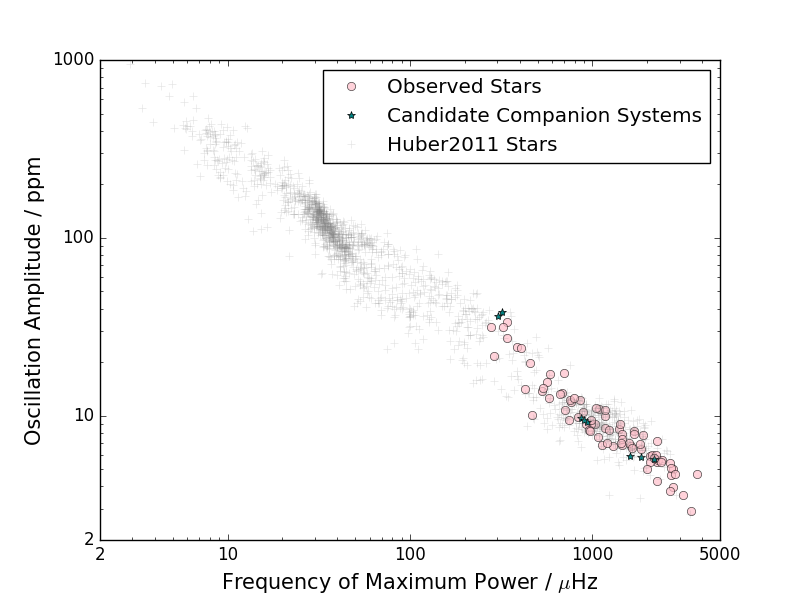}
\caption{Oscillation amplitude versus frequency of maximum power for all stars in \cite{Huber2011} (grey), all targets observed in our AO survey (pink circles) and all targets in the survey with detected secondary sources (green stars). Error bars omitted for clarity but are found in \cite{Huber2011}.}
\label{fig:AmpOscDia}
\end{figure}

\begin{figure}[]
\includegraphics[height = 7cm]{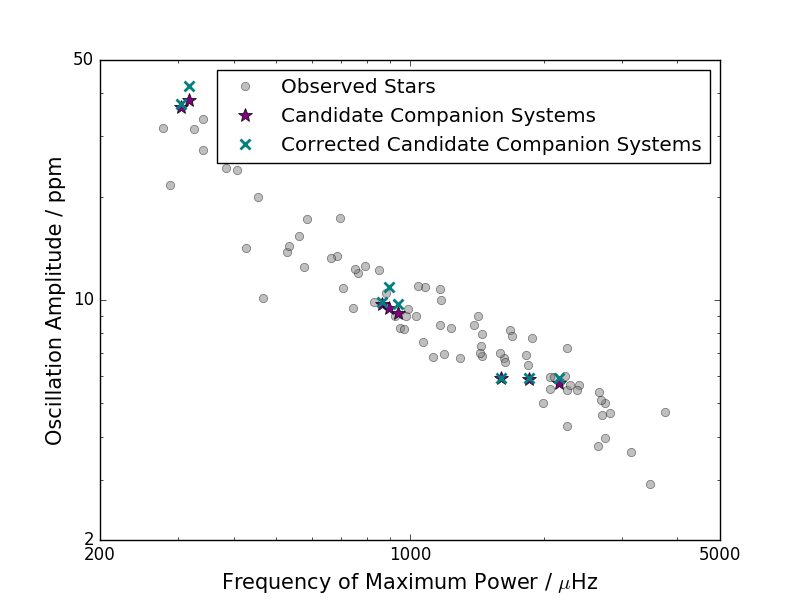}
\caption{Oscillation amplitude versus maximum oscillation frequency for targets in our sample. Candidate companion systems are shown before (purple stars) and after (green crosses) correction for the amplitude dilution from secondary sources.}
\label{fig:AmpOscDia2}
\end{figure}

The presence of undetected companions or secondary sources to these asteroseismic stars could provide an explanation for the additional scatter seen in the predictions of oscillation amplitude. These secondaries contribute additional radiant flux and hence cause a systematic dilution of the observed amplitudes in the \textit{Kepler} bandpass. To test this, we have used the asteroseismic data from \cite{Huber2011}, whose target sample contains all our candidate companion systems. Figure \ref{fig:AmpOscDia} shows the relation between oscillation amplitude and the frequency of maximum power ($\nu_{\rm max}$) for the full \citet{Huber2011} sample. For clarity, Figure \ref{fig:AmpOscDia2} shows only stars included in our survey and the oscillation amplitudes after correcting for the amplitude dilution by secondary sources. The correction of amplitude dilution has a negligible effect on the overall scatter of oscillation amplitudes vs. the frequency of maximum power. Fainter sources below our sensitivity limit will have amplitude dilutions of $<1$\% which will have an even smaller impact on the scatter. Therefore, the presence of stellar companions to asteroseismic stars is unlikely to be the sole cause of the large scatter. 


\subsection{Comparison of the Detected Companion Fraction to the Robo-AO \textit{Kepler} Planetary Candidate Survey}
\label{sec:CTORP}


We compared the candidate companion fraction from this, the asteroseismic sample to the Robo-AO \textit{Kepler} Planetary Candidate Survey sample (comprising \textgreater90\% of the KOI catalog, \citealt{Law2014, Baranec2016, Ziegler2016}) to determine if there is a fundamental difference between them.

The number of candidate companion systems divided by the size of the sample produces the candidate companion fractions. Errors on these fractions are determined using binomial statistics (e.g., \citealt{Burgasser2003}). While both surveys use Robo-AO and observe KICs, the asteroseismic sample is brighter, has a narrower range of effective temperatures and, on average, is observed in better conditions. 

Overall, for separations out to 2\farcs5, the asteroseismic sample has a lower companion fraction than the KOI sample and they do not agree within 1$\sigma$ error ($4.0\%\substack{+3.0\% \\ -1.2\%}$ and $8.7\%\substack{+0.5\% \\ -0.5\%}$ respectively). 


To determine a more comparable companion fraction for the KOI sample, we combined data from each of the KOI surveys and removed all stars with spectral types inconsistent with the asteroseismic sample. This meant removing all KOIs less than 3900K and above 7600K. 

Asteroseismic stars on average appear brighter than KOI stars, so we further restricted the KOI sample by primary star magnitude. Restricting the sample in the $i'$-band results in stars between 7.1 and 11.3 mag, making it consistent with the asteroseismic sample. When restricting by both temperature and magnitude the companion fraction for the KOI sample rose to $13.8\%\substack{+9.0\% \\ -5.8\%}$. We did not include stars without an observed $i'$-band magnitude.

This is unexpected as the asteroseismic sample survey comprises deeper contrast images on average and should therefore produce more candidate companions than the KOI survey. This discrepancy could be due to several factors, e.g., an astrophysical mechanism preventing asteroseismic stars from developing binary companions, small number statistics, or from a biased selection process for the the standard and/or KOI stars, due to small number statistics. We plan to survey the remaining standard stars list (N$\sim$400) in order to confirm if this discrepancy is real. 



\section{Conclusion}
\label{sec:C}

We used Robo-AO to observe 99 \textit{Kepler} stars that demonstrate stellar oscillations and found 8 candidate companion systems that dilute the oscillation amplitudes of their primary light curves. Amplitude dilution values amongst these stars range from 0.43\% to 15.4\% and does not explain the excess scatter in the relationship between asteroseismic oscillation amplitudes and the frequency of maximum power (Huber et al. 2011, Corsaro et al. 2013).

Using additional infrared photometry we calculated the photometric distances to the secondary sources of the candidate companion systems. We found that two of the secondary sources are likely foreground objects and at least six of the secondaries are background sources; however we cannot exclude the possibility that three of these may be physically associated. The measured companion fraction of our \textit{Kepler} asteroseismic sample is $4.0\%\substack{+3.0\% \\ -1.2\%}$\ (for separations out to 2\farcs5), and is lower than that found for KOIs. A larger sample of asteroseismic stars is needed to determine if this is an astrophysical, bias and/or small sample effect. 

We will use Robo-AO (now at the Kitt Peak 2.1-m telescope; \citealt{Salama2016}) to observe the remainder of the standard stars list, the Gold priority stars, to also determine whether these stars have binary companions or secondary sources within their \textit{Kepler} photometric apertures. 
 
\acknowledgments

This research is supported by the NASA Exoplanets Research Program, grant $\#$NNX 15AC91G. J.S. acknowledges the U.K. SLC. C.B. acknowledges support from the Alfred P. Sloan Foundation. We thank Joanna Bulger for helping with the Subaru observations.

The Robo-AO system was developed by collaborating partner institutions, the California Institute of Technology and the Inter-University Centre for Astronomy and Astrophysics, and with the support of the National Science Foundation under Grant Nos. AST-0906060, AST-0960343, and AST-1207891, the Mt. Cuba Astronomical Foundation and by a gift from Samuel Oschin. Some of the data presented herein were obtained at the W.M. Keck Observatory, which is operated as a scientific partnership among the California Institute of Technology, the University of California and the National Aeronautics and Space Administration. The Observatory was made possible by the generous financial support of the W.M. Keck Foundation. This work is based in part on data collected at Subaru Telescope, which is operated by the National Astronomical Observatory of Japan. SNIFS on the UH 2.2-m telescope is part of the Nearby Supernova Factory project, a scientific collaboration among the Centre de Recherche Astronomique de Lyon, Institut de Physique Nuclaire de Lyon, Laboratoire de Physique Nuclaire et des Hautes Energies, Lawrence Berkeley National Laboratory, Yale University, University of Bonn, Max Planck Institute for Astrophysics, Tsinghua Center for Astrophysics, and the Centre de Physique des Particules de Marseille. The authors wish to recognize and acknowledge the very significant cultural role and reverence that the summit of Maunakea has always had within the indigenous Hawaiian community. We are most fortunate to have the opportunity to conduct observations from this mountain.



{\it Facilities:} \facility{PO:1.5m (Robo-AO), Keck:II (NIRC2-NGS), Subaru (IRCS), UH 2.2m (SNIFS)}



\bibliographystyle{apj.bst}   
\bibliography{References.bib}  

\begin{thebibliography}{59}
\expandafter\ifx\csname natexlab\endcsname\relax\def\natexlab#1{#1}\fi

\bibitem[{{Adams} {et~al.}(2012){Adams}, {Ciardi}, {Dupree}, {Gautier},
  {Kulesa}, \& {McCarthy}}]{Adams2012}
{Adams}, E.~R., {Ciardi}, D.~R., {Dupree}, A.~K., {Gautier}, III, T.~N.,
  {Kulesa}, C., \& {McCarthy}, D. 2012, \aj, 144, 42

\bibitem[{{Aerts} {et~al.}(2010){Aerts}, {Christensen-Dalsgaard}, \&
  {Kurtz}}]{Aerts2010}
{Aerts}, C., {Christensen-Dalsgaard}, J., \& {Kurtz}, D.~W. 2010,
  {Asteroseismology} (Springer, Netherlands)

\bibitem[{{Aldering} {et~al.}(2002){Aldering}, {Adam}, {Antilogus}, {Astier},
  {Bacon}, {Bongard}, {Bonnaud}, {Copin}, {Hardin}, {Henault}, {Howell},
  {Lemonnier}, {Levy}, {Loken}, {Nugent}, {Pain}, {Pecontal}, {Pecontal},
  {Perlmutter}, {Quimby}, {Schahmaneche}, {Smadja}, \&
  {Wood-Vasey}}]{Aldering2002}
{Aldering}, G., {Adam}, G., {Antilogus}, P., {Astier}, P., {Bacon}, R.,
  {Bongard}, S., {Bonnaud}, C., {Copin}, Y., {Hardin}, D., {Henault}, F.,
  {Howell}, D.~A., {Lemonnier}, J.-P., {Levy}, J.-M., {Loken}, S.~C., {Nugent},
  P.~E., {Pain}, R., {Pecontal}, A., {Pecontal}, E., {Perlmutter}, S.,
  {Quimby}, R.~M., {Schahmaneche}, K., {Smadja}, G., \& {Wood-Vasey}, W.~M.
  2002, in \procspie, Vol. 4836, Survey and Other Telescope Technologies and
  Discoveries, ed. J.~A. {Tyson} \& S.~{Wolff}, 61--72

\bibitem[{{Appourchaux} {et~al.}(2014){Appourchaux}, {Antia}, {Benomar},
  {Campante}, {Davies}, {Handberg}, {Howe}, {R{\'e}gulo}, {Belkacem}, {Houdek},
  {Garc{\'{\i}}a}, \& {Chaplin}}]{Appourchaux2014}
{Appourchaux}, T., {Antia}, H.~M., {Benomar}, O., {Campante}, T.~L., {Davies},
  G.~R., {Handberg}, R., {Howe}, R., {R{\'e}gulo}, C., {Belkacem}, K.,
  {Houdek}, G., {Garc{\'{\i}}a}, R.~A., \& {Chaplin}, W.~J. 2014, \aap, 566,
  A20

\bibitem[{{Atkinson} {et~al.}(2017){Atkinson}, {Baranec}, {Ziegler}, {Law}, \&
  {Morton}}]{Atkinson2016}
{Atkinson}, D., {Baranec}, C., {Ziegler}, C., {Law}, N., \& {Morton}, T. 2017,
  \aj, 153, 25

\bibitem[{{Auvergne} {et~al.}(2009){Auvergne}, {Bodin}, {Boisnard}, {Buey},
  {Chaintreuil}, {Epstein}, {Jouret}, {Lam-Trong}, {Levacher}, {Magnan},
  {Perez}, {Plasson}, {Plesseria}, {Peter}, {Steller}, {Tiph{\`e}ne}, {Baglin},
  {Agogu{\'e}}, {Appourchaux}, {Barbet}, {Beaufort}, {Bellenger}, {Berlin},
  {Bernardi}, {Blouin}, {Boumier}, {Bonneau}, {Briet}, {Butler}, {Cautain},
  {Chiavassa}, {Costes}, {Cuvilho}, {Cunha-Parro}, {de Oliveira Fialho},
  {Decaudin}, {Defise}, {Djalal}, {Docclo}, {Drummond}, {Dupuis}, {Exil},
  {Faur{\'e}}, {Gaboriaud}, {Gamet}, {Gavalda}, {Grolleau}, {Gueguen},
  {Guivarc'h}, {Guterman}, {Hasiba}, {Huntzinger}, {Hustaix}, {Imbert},
  {Jeanville}, {Johlander}, {Jorda}, {Journoud}, {Karioty}, {Kerjean},
  {Lafond}, {Lapeyrere}, {Landiech}, {Larqu{\'e}}, {Laudet}, {Le Merrer},
  {Leporati}, {Leruyet}, {Levieuge}, {Llebaria}, {Martin}, {Mazy}, {Mesnager},
  {Michel}, {Moalic}, {Monjoin}, {Naudet}, {Neukirchner}, {Nguyen-Kim},
  {Ollivier}, {Orcesi}, {Ottacher}, {Oulali}, {Parisot}, {Perruchot},
  {Piacentino}, {Pinheiro da Silva}, {Platzer}, {Pontet}, {Pradines},
  {Quentin}, {Rohbeck}, {Rolland}, {Rollenhagen}, {Romagnan}, {Russ}, {Samadi},
  {Schmidt}, {Schwartz}, {Sebbag}, {Smit}, {Sunter}, {Tello}, {Toulouse},
  {Ulmer}, {Vandermarcq}, {Vergnault}, {Wallner}, {Waultier}, \&
  {Zanatta}}]{Auvergne2009}
{Auvergne}, M., {Bodin}, P., {Boisnard}, L., {Buey}, J.-T., {Chaintreuil}, S.,
  {Epstein}, G., {Jouret}, M., {Lam-Trong}, T., {Levacher}, P., {Magnan}, A.,
  {Perez}, R., {Plasson}, P., {Plesseria}, J., {Peter}, G., {Steller}, M.,
  {Tiph{\`e}ne}, D., {Baglin}, A., {Agogu{\'e}}, P., {Appourchaux}, T.,
  {Barbet}, D., {Beaufort}, T., {Bellenger}, R., {Berlin}, R., {Bernardi}, P.,
  {Blouin}, D., {Boumier}, P., {Bonneau}, F., {Briet}, R., {Butler}, B.,
  {Cautain}, R., {Chiavassa}, F., {Costes}, V., {Cuvilho}, J., {Cunha-Parro},
  V., {de Oliveira Fialho}, F., {Decaudin}, M., {Defise}, J.-M., {Djalal}, S.,
  {Docclo}, A., {Drummond}, R., {Dupuis}, O., {Exil}, G., {Faur{\'e}}, C.,
  {Gaboriaud}, A., {Gamet}, P., {Gavalda}, P., {Grolleau}, E., {Gueguen}, L.,
  {Guivarc'h}, V., {Guterman}, P., {Hasiba}, J., {Huntzinger}, G., {Hustaix},
  H., {Imbert}, C., {Jeanville}, G., {Johlander}, B., {Jorda}, L., {Journoud},
  P., {Karioty}, F., {Kerjean}, L., {Lafond}, L., {Lapeyrere}, V., {Landiech},
  P., {Larqu{\'e}}, T., {Laudet}, P., {Le Merrer}, J., {Leporati}, L.,
  {Leruyet}, B., {Levieuge}, B., {Llebaria}, A., {Martin}, L., {Mazy}, E.,
  {Mesnager}, J.-M., {Michel}, J.-P., {Moalic}, J.-P., {Monjoin}, W., {Naudet},
  D., {Neukirchner}, S., {Nguyen-Kim}, K., {Ollivier}, M., {Orcesi}, J.-L.,
  {Ottacher}, H., {Oulali}, A., {Parisot}, J., {Perruchot}, S., {Piacentino},
  A., {Pinheiro da Silva}, L., {Platzer}, J., {Pontet}, B., {Pradines}, A.,
  {Quentin}, C., {Rohbeck}, U., {Rolland}, G., {Rollenhagen}, F., {Romagnan},
  R., {Russ}, N., {Samadi}, R., {Schmidt}, R., {Schwartz}, N., {Sebbag}, I.,
  {Smit}, H., {Sunter}, W., {Tello}, M., {Toulouse}, P., {Ulmer}, B.,
  {Vandermarcq}, O., {Vergnault}, E., {Wallner}, R., {Waultier}, G., \&
  {Zanatta}, P. 2009, \aap, 506, 411

\bibitem[{{Baranec} {et~al.}(2014){Baranec}, {Riddle}, {Law}, {Ramaprakash},
  {Tendulkar}, {Hogstrom}, {Bui}, {Burse}, {Chordia}, {Das}, {Dekany},
  {Kulkarni}, \& {Punnadi}}]{Baranec2014}
{Baranec}, C., {Riddle}, R., {Law}, N.~M., {Ramaprakash}, A.~N., {Tendulkar},
  S., {Hogstrom}, K., {Bui}, K., {Burse}, M., {Chordia}, P., {Das}, H.,
  {Dekany}, R., {Kulkarni}, S., \& {Punnadi}, S. 2014, \apjl, 790, L8

\bibitem[{{Baranec} {et~al.}(2013){Baranec}, {Riddle}, {Law}, {Ramaprakash},
  {Tendulkar}, {Bui}, {Burse}, {Chordia}, {Das}, {Davis}, {Dekany}, {Kasliwal},
  {Kulkarni}, {Morton}, {Ofek}, \& {Punnadi}}]{Baranec2013}
{Baranec}, C., {Riddle}, R., {Law}, N.~M., {Ramaprakash}, A.~N., {Tendulkar},
  S.~P., {Bui}, K., {Burse}, M.~P., {Chordia}, P., {Das}, H.~K., {Davis},
  J.~T.~C., {Dekany}, R.~G., {Kasliwal}, M.~M., {Kulkarni}, S.~R., {Morton},
  T.~D., {Ofek}, E.~O., \& {Punnadi}, S. 2013, Journal of Visualized
  Experiments, 72

\bibitem[{{Baranec} {et~al.}(2016){Baranec}, {Ziegler}, {Law}, {Morton},
  {Riddle}, {Atkinson}, {Schonhut}, \& {Crepp}}]{Baranec2016}
{Baranec}, C., {Ziegler}, C., {Law}, N.~M., {Morton}, T., {Riddle}, R.,
  {Atkinson}, D., {Schonhut}, J., \& {Crepp}, J. 2016, \aj, 152, 18

\bibitem[{{Bedding} {et~al.}(2010){Bedding}, {Huber}, {Stello}, {Elsworth},
  {Hekker}, {Kallinger}, {Mathur}, {Mosser}, {Preston}, {Ballot}, {Barban},
  {Broomhall}, {Buzasi}, {Chaplin}, {Garc{\'{\i}}a}, {Gruberbauer}, {Hale}, {De
  Ridder}, {Frandsen}, {Borucki}, {Brown}, {Christensen-Dalsgaard},
  {Gilliland}, {Jenkins}, {Kjeldsen}, {Koch}, {Belkacem}, {Bildsten}, {Bruntt},
  {Campante}, {Deheuvels}, {Derekas}, {Dupret}, {Goupil}, {Hatzes}, {Houdek},
  {Ireland}, {Jiang}, {Karoff}, {Kiss}, {Lebreton}, {Miglio}, {Montalb{\'a}n},
  {Noels}, {Roxburgh}, {Sangaralingam}, {Stevens}, {Suran}, {Tarrant}, \&
  {Weiss}}]{Bedding2010}
{Bedding}, T.~R., {Huber}, D., {Stello}, D., {Elsworth}, Y.~P., {Hekker}, S.,
  {Kallinger}, T., {Mathur}, S., {Mosser}, B., {Preston}, H.~L., {Ballot}, J.,
  {Barban}, C., {Broomhall}, A.~M., {Buzasi}, D.~L., {Chaplin}, W.~J.,
  {Garc{\'{\i}}a}, R.~A., {Gruberbauer}, M., {Hale}, S.~J., {De Ridder}, J.,
  {Frandsen}, S., {Borucki}, W.~J., {Brown}, T., {Christensen-Dalsgaard}, J.,
  {Gilliland}, R.~L., {Jenkins}, J.~M., {Kjeldsen}, H., {Koch}, D., {Belkacem},
  K., {Bildsten}, L., {Bruntt}, H., {Campante}, T.~L., {Deheuvels}, S.,
  {Derekas}, A., {Dupret}, M.-A., {Goupil}, M.-J., {Hatzes}, A., {Houdek}, G.,
  {Ireland}, M.~J., {Jiang}, C., {Karoff}, C., {Kiss}, L.~L., {Lebreton}, Y.,
  {Miglio}, A., {Montalb{\'a}n}, J., {Noels}, A., {Roxburgh}, I.~W.,
  {Sangaralingam}, V., {Stevens}, I.~R., {Suran}, M.~D., {Tarrant}, N.~J., \&
  {Weiss}, A. 2010, \apjl, 713, L176

\bibitem[{{Bedding} {et~al.}(2011){Bedding}, {Mosser}, {Huber},
  {Montalb{\'a}n}, {Beck}, {Christensen-Dalsgaard}, {Elsworth},
  {Garc{\'{\i}}a}, {Miglio}, {Stello}, {White}, {De Ridder}, {Hekker}, {Aerts},
  {Barban}, {Belkacem}, {Broomhall}, {Brown}, {Buzasi}, {Carrier}, {Chaplin},
  {di Mauro}, {Dupret}, {Frandsen}, {Gilliland}, {Goupil}, {Jenkins},
  {Kallinger}, {Kawaler}, {Kjeldsen}, {Mathur}, {Noels}, {Silva Aguirre}, \&
  {Ventura}}]{Bedding2011}
{Bedding}, T.~R., {Mosser}, B., {Huber}, D., {Montalb{\'a}n}, J., {Beck}, P.,
  {Christensen-Dalsgaard}, J., {Elsworth}, Y.~P., {Garc{\'{\i}}a}, R.~A.,
  {Miglio}, A., {Stello}, D., {White}, T.~R., {De Ridder}, J., {Hekker}, S.,
  {Aerts}, C., {Barban}, C., {Belkacem}, K., {Broomhall}, A.-M., {Brown},
  T.~M., {Buzasi}, D.~L., {Carrier}, F., {Chaplin}, W.~J., {di Mauro}, M.~P.,
  {Dupret}, M.-A., {Frandsen}, S., {Gilliland}, R.~L., {Goupil}, M.-J.,
  {Jenkins}, J.~M., {Kallinger}, T., {Kawaler}, S., {Kjeldsen}, H., {Mathur},
  S., {Noels}, A., {Silva Aguirre}, V., \& {Ventura}, P. 2011, \nat, 471, 608

\bibitem[{{Borucki} {et~al.}(2010){Borucki}, {Koch}, {Basri}, {Batalha},
  {Brown}, {Caldwell}, {Caldwell}, {Christensen-Dalsgaard}, {Cochran},
  {DeVore}, {Dunham}, {Dupree}, {Gautier}, {Geary}, {Gilliland}, {Gould},
  {Howell}, {Jenkins}, {Kondo}, {Latham}, {Marcy}, {Meibom}, {Kjeldsen},
  {Lissauer}, {Monet}, {Morrison}, {Sasselov}, {Tarter}, {Boss}, {Brownlee},
  {Owen}, {Buzasi}, {Charbonneau}, {Doyle}, {Fortney}, {Ford}, {Holman},
  {Seager}, {Steffen}, {Welsh}, {Rowe}, {Anderson}, {Buchhave}, {Ciardi},
  {Walkowicz}, {Sherry}, {Horch}, {Isaacson}, {Everett}, {Fischer}, {Torres},
  {Johnson}, {Endl}, {MacQueen}, {Bryson}, {Dotson}, {Haas}, {Kolodziejczak},
  {Van Cleve}, {Chandrasekaran}, {Twicken}, {Quintana}, {Clarke}, {Allen},
  {Li}, {Wu}, {Tenenbaum}, {Verner}, {Bruhweiler}, {Barnes}, \&
  {Prsa}}]{Borucki2010}
{Borucki}, W.~J., {Koch}, D., {Basri}, G., {Batalha}, N., {Brown}, T.,
  {Caldwell}, D., {Caldwell}, J., {Christensen-Dalsgaard}, J., {Cochran},
  W.~D., {DeVore}, E., {Dunham}, E.~W., {Dupree}, A.~K., {Gautier}, T.~N.,
  {Geary}, J.~C., {Gilliland}, R., {Gould}, A., {Howell}, S.~B., {Jenkins},
  J.~M., {Kondo}, Y., {Latham}, D.~W., {Marcy}, G.~W., {Meibom}, S.,
  {Kjeldsen}, H., {Lissauer}, J.~J., {Monet}, D.~G., {Morrison}, D.,
  {Sasselov}, D., {Tarter}, J., {Boss}, A., {Brownlee}, D., {Owen}, T.,
  {Buzasi}, D., {Charbonneau}, D., {Doyle}, L., {Fortney}, J., {Ford}, E.~B.,
  {Holman}, M.~J., {Seager}, S., {Steffen}, J.~H., {Welsh}, W.~F., {Rowe}, J.,
  {Anderson}, H., {Buchhave}, L., {Ciardi}, D., {Walkowicz}, L., {Sherry}, W.,
  {Horch}, E., {Isaacson}, H., {Everett}, M.~E., {Fischer}, D., {Torres}, G.,
  {Johnson}, J.~A., {Endl}, M., {MacQueen}, P., {Bryson}, S.~T., {Dotson}, J.,
  {Haas}, M., {Kolodziejczak}, J., {Van Cleve}, J., {Chandrasekaran}, H.,
  {Twicken}, J.~D., {Quintana}, E.~V., {Clarke}, B.~D., {Allen}, C., {Li}, J.,
  {Wu}, H., {Tenenbaum}, P., {Verner}, E., {Bruhweiler}, F., {Barnes}, J., \&
  {Prsa}, A. 2010, Science, 327, 977

\bibitem[{{Buton} {et~al.}(2013){Buton}, {Copin}, {Aldering}, {Antilogus},
  {Aragon}, {Bailey}, {Baltay}, {Bongard}, {Canto}, {Cellier-Holzem},
  {Childress}, {Chotard}, {Fakhouri}, {Gangler}, {Guy}, {Hsiao}, {Kerschhaggl},
  {Kowalski}, {Loken}, {Nugent}, {Paech}, {Pain}, {P{\'e}contal}, {Pereira},
  {Perlmutter}, {Rabinowitz}, {Rigault}, {Runge}, {Scalzo}, {Smadja}, {Tao},
  {Thomas}, {Weaver}, {Wu}, \& {Nearby SuperNova Factory}}]{Buton2012}
{Buton}, C., {Copin}, Y., {Aldering}, G., {Antilogus}, P., {Aragon}, C.,
  {Bailey}, S., {Baltay}, C., {Bongard}, S., {Canto}, A., {Cellier-Holzem}, F.,
  {Childress}, M., {Chotard}, N., {Fakhouri}, H.~K., {Gangler}, E., {Guy}, J.,
  {Hsiao}, E.~Y., {Kerschhaggl}, M., {Kowalski}, M., {Loken}, S., {Nugent}, P.,
  {Paech}, K., {Pain}, R., {P{\'e}contal}, E., {Pereira}, R., {Perlmutter}, S.,
  {Rabinowitz}, D., {Rigault}, M., {Runge}, K., {Scalzo}, R., {Smadja}, G.,
  {Tao}, C., {Thomas}, R.~C., {Weaver}, B.~A., {Wu}, C., \& {Nearby SuperNova
  Factory}. 2013, \aap, 549, A8

\bibitem[{{Campante} {et~al.}(2015){Campante}, {Barclay}, {Swift}, {Huber},
  {Adibekyan}, {Cochran}, {Burke}, {Isaacson}, {Quintana}, {Davies}, {Silva
  Aguirre}, {Ragozzine}, {Riddle}, {Baranec}, {Basu}, {Chaplin},
  {Christensen-Dalsgaard}, {Metcalfe}, {Bedding}, {Handberg}, {Stello},
  {Brewer}, {Hekker}, {Karoff}, {Kolbl}, {Law}, {Lundkvist}, {Miglio}, {Rowe},
  {Santos}, {Van Laerhoven}, {Arentoft}, {Elsworth}, {Fischer}, {Kawaler},
  {Kjeldsen}, {Lund}, {Marcy}, {Sousa}, {Sozzetti}, \& {White}}]{Campante2015}
{Campante}, T.~L., {Barclay}, T., {Swift}, J.~J., {Huber}, D., {Adibekyan},
  V.~Z., {Cochran}, W., {Burke}, C.~J., {Isaacson}, H., {Quintana}, E.~V.,
  {Davies}, G.~R., {Silva Aguirre}, V., {Ragozzine}, D., {Riddle}, R.,
  {Baranec}, C., {Basu}, S., {Chaplin}, W.~J., {Christensen-Dalsgaard}, J.,
  {Metcalfe}, T.~S., {Bedding}, T.~R., {Handberg}, R., {Stello}, D., {Brewer},
  J.~M., {Hekker}, S., {Karoff}, C., {Kolbl}, R., {Law}, N.~M., {Lundkvist},
  M., {Miglio}, A., {Rowe}, J.~F., {Santos}, N.~C., {Van Laerhoven}, C.,
  {Arentoft}, T., {Elsworth}, Y.~P., {Fischer}, D.~A., {Kawaler}, S.~D.,
  {Kjeldsen}, H., {Lund}, M.~N., {Marcy}, G.~W., {Sousa}, S.~G., {Sozzetti},
  A., \& {White}, T.~R. 2015, \apj, 799, 170

\bibitem[{{Campante} {et~al.}(2014){Campante}, {Chaplin}, {Lund}, {Huber},
  {Hekker}, {Garc{\'{\i}}a}, {Corsaro}, {Handberg}, {Miglio}, {Arentoft},
  {Basu}, {Bedding}, {Christensen-Dalsgaard}, {Davies}, {Elsworth},
  {Gilliland}, {Karoff}, {Kawaler}, {Kjeldsen}, {Lundkvist}, {Metcalfe}, {Silva
  Aguirre}, \& {Stello}}]{Campante2014}
{Campante}, T.~L., {Chaplin}, W.~J., {Lund}, M.~N., {Huber}, D., {Hekker}, S.,
  {Garc{\'{\i}}a}, R.~A., {Corsaro}, E., {Handberg}, R., {Miglio}, A.,
  {Arentoft}, T., {Basu}, S., {Bedding}, T.~R., {Christensen-Dalsgaard}, J.,
  {Davies}, G.~R., {Elsworth}, Y.~P., {Gilliland}, R.~L., {Karoff}, C.,
  {Kawaler}, S.~D., {Kjeldsen}, H., {Lundkvist}, M., {Metcalfe}, T.~S., {Silva
  Aguirre}, V., \& {Stello}, D. 2014, \apj, 783, 123

\bibitem[{{Cardelli} {et~al.}(1989){Cardelli}, {Clayton}, \&
  {Mathis}}]{Cardelli1989}
{Cardelli}, J.~A., {Clayton}, G.~C., \& {Mathis}, J.~S. 1989, \apj, 345, 245

\bibitem[{{Cenko} {et~al.}(2006){Cenko}, {Fox}, {Moon}, {Harrison}, {Kulkarni},
  {Henning}, {Guzman}, {Bonati}, {Smith}, {Thicksten}, {Doyle}, {Petrie},
  {Gal-Yam}, {Soderberg}, {Anagnostou}, \& {Laity}}]{Cenko2006}
{Cenko}, S.~B., {Fox}, D.~B., {Moon}, D.~S., {Harrison}, F.~A., {Kulkarni},
  S.~R., {Henning}, J.~R., {Guzman}, C.~D., {Bonati}, M., {Smith}, R.~M.,
  {Thicksten}, R.~P., {Doyle}, M.~W., {Petrie}, H.~L., {Gal-Yam}, A.,
  {Soderberg}, A.~M., {Anagnostou}, N.~L., \& {Laity}, A.~C. 2006, \pasp, 118,
  1396

\bibitem[{{Chaplin} {et~al.}(2010){Chaplin}, {Appourchaux}, {Elsworth},
  {Garc{\'{\i}}a}, {Houdek}, {Karoff}, {Metcalfe}, {Molenda-{\.Z}akowicz},
  {Monteiro}, {Thompson}, {Brown}, {Christensen-Dalsgaard}, {Gilliland},
  {Kjeldsen}, {Borucki}, {Koch}, {Jenkins}, {Ballot}, {Basu}, {Bazot},
  {Bedding}, {Benomar}, {Bonanno}, {Brand{\~a}o}, {Bruntt}, {Campante},
  {Creevey}, {Di Mauro}, {Do{\v g}an}, {Dreizler}, {Eggenberger}, {Esch},
  {Fletcher}, {Frandsen}, {Gai}, {Gaulme}, {Handberg}, {Hekker}, {Howe},
  {Huber}, {Korzennik}, {Lebrun}, {Leccia}, {Martic}, {Mathur}, {Mosser},
  {New}, {Quirion}, {R{\'e}gulo}, {Roxburgh}, {Salabert}, {Schou}, {Sousa},
  {Stello}, {Verner}, {Arentoft}, {Barban}, {Belkacem}, {Benatti}, {Biazzo},
  {Boumier}, {Bradley}, {Broomhall}, {Buzasi}, {Claudi}, {Cunha}, {D'Antona},
  {Deheuvels}, {Derekas}, {Garc{\'{\i}}a Hern{\'a}ndez}, {Giampapa}, {Goupil},
  {Gruberbauer}, {Guzik}, {Hale}, {Ireland}, {Kiss}, {Kitiashvili},
  {Kolenberg}, {Korhonen}, {Kosovichev}, {Kupka}, {Lebreton}, {Leroy},
  {Ludwig}, {Mathis}, {Michel}, {Miglio}, {Montalb{\'a}n}, {Moya}, {Noels},
  {Noyes}, {Pall{\'e}}, {Piau}, {Preston}, {Roca Cort{\'e}s}, {Roth}, {Sato},
  {Schmitt}, {Serenelli}, {Silva Aguirre}, {Stevens}, {Su{\'a}rez}, {Suran},
  {Trampedach}, {Turck-Chi{\`e}ze}, {Uytterhoeven}, {Ventura}, \&
  {Wilson}}]{Chaplin2010}
{Chaplin}, W.~J., {Appourchaux}, T., {Elsworth}, Y., {Garc{\'{\i}}a}, R.~A.,
  {Houdek}, G., {Karoff}, C., {Metcalfe}, T.~S., {Molenda-{\.Z}akowicz}, J.,
  {Monteiro}, M.~J.~P.~F.~G., {Thompson}, M.~J., {Brown}, T.~M.,
  {Christensen-Dalsgaard}, J., {Gilliland}, R.~L., {Kjeldsen}, H., {Borucki},
  W.~J., {Koch}, D., {Jenkins}, J.~M., {Ballot}, J., {Basu}, S., {Bazot}, M.,
  {Bedding}, T.~R., {Benomar}, O., {Bonanno}, A., {Brand{\~a}o}, I.~M.,
  {Bruntt}, H., {Campante}, T.~L., {Creevey}, O.~L., {Di Mauro}, M.~P., {Do{\v
  g}an}, G., {Dreizler}, S., {Eggenberger}, P., {Esch}, L., {Fletcher}, S.~T.,
  {Frandsen}, S., {Gai}, N., {Gaulme}, P., {Handberg}, R., {Hekker}, S.,
  {Howe}, R., {Huber}, D., {Korzennik}, S.~G., {Lebrun}, J.~C., {Leccia}, S.,
  {Martic}, M., {Mathur}, S., {Mosser}, B., {New}, R., {Quirion}, P.-O.,
  {R{\'e}gulo}, C., {Roxburgh}, I.~W., {Salabert}, D., {Schou}, J., {Sousa},
  S.~G., {Stello}, D., {Verner}, G.~A., {Arentoft}, T., {Barban}, C.,
  {Belkacem}, K., {Benatti}, S., {Biazzo}, K., {Boumier}, P., {Bradley}, P.~A.,
  {Broomhall}, A.-M., {Buzasi}, D.~L., {Claudi}, R.~U., {Cunha}, M.~S.,
  {D'Antona}, F., {Deheuvels}, S., {Derekas}, A., {Garc{\'{\i}}a
  Hern{\'a}ndez}, A., {Giampapa}, M.~S., {Goupil}, M.~J., {Gruberbauer}, M.,
  {Guzik}, J.~A., {Hale}, S.~J., {Ireland}, M.~J., {Kiss}, L.~L.,
  {Kitiashvili}, I.~N., {Kolenberg}, K., {Korhonen}, H., {Kosovichev}, A.~G.,
  {Kupka}, F., {Lebreton}, Y., {Leroy}, B., {Ludwig}, H.-G., {Mathis}, S.,
  {Michel}, E., {Miglio}, A., {Montalb{\'a}n}, J., {Moya}, A., {Noels}, A.,
  {Noyes}, R.~W., {Pall{\'e}}, P.~L., {Piau}, L., {Preston}, H.~L., {Roca
  Cort{\'e}s}, T., {Roth}, M., {Sato}, K.~H., {Schmitt}, J., {Serenelli},
  A.~M., {Silva Aguirre}, V., {Stevens}, I.~R., {Su{\'a}rez}, J.~C., {Suran},
  M.~D., {Trampedach}, R., {Turck-Chi{\`e}ze}, S., {Uytterhoeven}, K.,
  {Ventura}, R., \& {Wilson}, P.~A. 2010, \apjl, 713, L169

\bibitem[{{Chaplin} {et~al.}(2014){Chaplin}, {Basu}, {Huber}, {Serenelli},
  {Casagrande}, {Silva Aguirre}, {Ball}, {Creevey}, {Gizon}, {Handberg},
  {Karoff}, {Lutz}, {Marques}, {Miglio}, {Stello}, {Suran}, {Pricopi},
  {Metcalfe}, {Monteiro}, {Molenda-{\.Z}akowicz}, {Appourchaux},
  {Christensen-Dalsgaard}, {Elsworth}, {Garc{\'{\i}}a}, {Houdek}, {Kjeldsen},
  {Bonanno}, {Campante}, {Corsaro}, {Gaulme}, {Hekker}, {Mathur}, {Mosser},
  {R{\'e}gulo}, \& {Salabert}}]{Chaplin2014}
{Chaplin}, W.~J., {Basu}, S., {Huber}, D., {Serenelli}, A., {Casagrande}, L.,
  {Silva Aguirre}, V., {Ball}, W.~H., {Creevey}, O.~L., {Gizon}, L.,
  {Handberg}, R., {Karoff}, C., {Lutz}, R., {Marques}, J.~P., {Miglio}, A.,
  {Stello}, D., {Suran}, M.~D., {Pricopi}, D., {Metcalfe}, T.~S., {Monteiro},
  M.~J.~P.~F.~G., {Molenda-{\.Z}akowicz}, J., {Appourchaux}, T.,
  {Christensen-Dalsgaard}, J., {Elsworth}, Y., {Garc{\'{\i}}a}, R.~A.,
  {Houdek}, G., {Kjeldsen}, H., {Bonanno}, A., {Campante}, T.~L., {Corsaro},
  E., {Gaulme}, P., {Hekker}, S., {Mathur}, S., {Mosser}, B., {R{\'e}gulo}, C.,
  \& {Salabert}, D. 2014, \apjs, 210, 1

\bibitem[{{Christensen-Dalsgaard}(2014)}]{Christensen-Dalsgaard2014}
{Christensen-Dalsgaard}, J. 2014, Danmarks Grundforskningsfond: Institut for
  Fysik og Astronomi, Aarhus Universitet Teoretisk Astrofysik Center.

\bibitem[{{Corsaro} {et~al.}(2013){Corsaro}, {Fr{\"o}hlich}, {Bonanno},
  {Huber}, {Bedding}, {Benomar}, {De Ridder}, \& {Stello}}]{corsaro13}
{Corsaro}, E., {Fr{\"o}hlich}, H.-E., {Bonanno}, A., {Huber}, D., {Bedding},
  T.~R., {Benomar}, O., {De Ridder}, J., \& {Stello}, D. 2013, \mnras, 430,
  2313

\bibitem[{{Davies} {et~al.}(2016){Davies}, {Silva Aguirre}, {Bedding},
  {Handberg}, {Lund}, {Chaplin}, {Huber}, {White}, {Benomar}, {Hekker}, {Basu},
  {Campante}, {Christensen-Dalsgaard}, {Elsworth}, {Karoff}, {Kjeldsen},
  {Lundkvist}, {Metcalfe}, \& {Stello}}]{Davies2016}
{Davies}, G.~R., {Silva Aguirre}, V., {Bedding}, T.~R., {Handberg}, R., {Lund},
  M.~N., {Chaplin}, W.~J., {Huber}, D., {White}, T.~R., {Benomar}, O.,
  {Hekker}, S., {Basu}, S., {Campante}, T.~L., {Christensen-Dalsgaard}, J.,
  {Elsworth}, Y., {Karoff}, C., {Kjeldsen}, H., {Lundkvist}, M.~S., {Metcalfe},
  T.~S., \& {Stello}, D. 2016, \mnras, 456, 2183

\bibitem[{{Deheuvels} {et~al.}(2012){Deheuvels}, {Garc{\'{\i}}a}, {Chaplin},
  {Basu}, {Antia}, {Appourchaux}, {Benomar}, {Davies}, {Elsworth}, {Gizon},
  {Goupil}, {Reese}, {Regulo}, {Schou}, {Stahn}, {Casagrande},
  {Christensen-Dalsgaard}, {Fischer}, {Hekker}, {Kjeldsen}, {Mathur}, {Mosser},
  {Pinsonneault}, {Valenti}, {Christiansen}, {Kinemuchi}, \&
  {Mullally}}]{Deheuvels2012}
{Deheuvels}, S., {Garc{\'{\i}}a}, R.~A., {Chaplin}, W.~J., {Basu}, S., {Antia},
  H.~M., {Appourchaux}, T., {Benomar}, O., {Davies}, G.~R., {Elsworth}, Y.,
  {Gizon}, L., {Goupil}, M.~J., {Reese}, D.~R., {Regulo}, C., {Schou}, J.,
  {Stahn}, T., {Casagrande}, L., {Christensen-Dalsgaard}, J., {Fischer}, D.,
  {Hekker}, S., {Kjeldsen}, H., {Mathur}, S., {Mosser}, B., {Pinsonneault}, M.,
  {Valenti}, J., {Christiansen}, J.~L., {Kinemuchi}, K., \& {Mullally}, F.
  2012, \apj, 756, 19

\bibitem[{{Dupuy} {et~al.}(2016){Dupuy}, {Kratter}, {Kraus}, {Isaacson},
  {Mann}, {Ireland}, {Howard}, \& {Huber}}]{Dupuy2016}
{Dupuy}, T.~J., {Kratter}, K.~M., {Kraus}, A.~L., {Isaacson}, H., {Mann},
  A.~W., {Ireland}, M.~J., {Howard}, A.~W., \& {Huber}, D. 2016, \apj, 817, 80

\bibitem[{{Garc{\'{\i}}a} {et~al.}(2014){Garc{\'{\i}}a}, {Mathur}, {Pires},
  {R{\'e}gulo}, {Bellamy}, {Pall{\'e}}, {Ballot}, {Barcel{\'o} Forteza},
  {Beck}, {Bedding}, {Ceillier}, {Roca Cort{\'e}s}, {Salabert}, \&
  {Stello}}]{Garcia2014}
{Garc{\'{\i}}a}, R.~A., {Mathur}, S., {Pires}, S., {R{\'e}gulo}, C., {Bellamy},
  B., {Pall{\'e}}, P.~L., {Ballot}, J., {Barcel{\'o} Forteza}, S., {Beck},
  P.~G., {Bedding}, T.~R., {Ceillier}, T., {Roca Cort{\'e}s}, T., {Salabert},
  D., \& {Stello}, D. 2014, \aap, 568, A10

\bibitem[{{Gilliland} {et~al.}(2010){Gilliland}, {Jenkins}, {Borucki},
  {Bryson}, {Caldwell}, {Clarke}, {Dotson}, {Haas}, {Hall}, {Klaus}, {Koch},
  {McCauliff}, {Quintana}, {Twicken}, \& {van Cleve}}]{Gilliland2010}
{Gilliland}, R.~L., {Jenkins}, J.~M., {Borucki}, W.~J., {Bryson}, S.~T.,
  {Caldwell}, D.~A., {Clarke}, B.~D., {Dotson}, J.~L., {Haas}, M.~R., {Hall},
  J., {Klaus}, T., {Koch}, D., {McCauliff}, S., {Quintana}, E.~V., {Twicken},
  J.~D., \& {van Cleve}, J.~E. 2010, \apjl, 713, L160

\bibitem[{{Ginski} {et~al.}(2016){Ginski}, {Mugrauer}, {Seeliger}, {Buder},
  {Errmann}, {Avenhaus}, {Mouillet}, {Maire}, \& {Raetz}}]{Ginski2016}
{Ginski}, C., {Mugrauer}, M., {Seeliger}, M., {Buder}, S., {Errmann}, R.,
  {Avenhaus}, H., {Mouillet}, D., {Maire}, A.-L., \& {Raetz}, S. 2016, \mnras,
  457, 2173

\bibitem[{{Hacking} {et~al.}(1997){Hacking}, {Herter}, {Stacey}, {Houck},
  {Shupe}, {Lonsdale}, {Gautier}, {Schember}, {Werner}, {Soifer}, {Moseley}, \&
  {Graf}}]{Hacking1997}
{Hacking}, P., {Herter}, T., {Stacey}, C., {Houck}, J.~R., {Shupe}, D.~L.,
  {Lonsdale}, C., {Gautier}, T.~N., {Schember}, H.~R., {Werner}, M.~W.,
  {Soifer}, B.~T., {Moseley}, S.~H., \& {Graf}, P. 1997, in Astronomical
  Society of the Pacific Conference Series, Vol. 124, Diffuse Infrared
  Radiation and the IRTS, ed. H.~{Okuda}, T.~{Matsumoto}, \& T.~{Rollig}, 432

\bibitem[{{Handler}(2013)}]{Handler2013}
{Handler}, G. {Asteroseismology} (Handler), 207

\bibitem[{{Hekker} {et~al.}(2011){Hekker}, {Gilliland}, {Elsworth}, {Chaplin},
  {De Ridder}, {Stello}, {Kallinger}, {Ibrahim}, {Klaus}, \& {Li}}]{Hekker2011}
{Hekker}, S., {Gilliland}, R.~L., {Elsworth}, Y., {Chaplin}, W.~J., {De
  Ridder}, J., {Stello}, D., {Kallinger}, T., {Ibrahim}, K.~A., {Klaus}, T.~C.,
  \& {Li}, J. 2011, \mnras, 414, 2594

\bibitem[{Holwards(1642)}]{Holwards1642}
Holwards, J. 1642, Epitome astronomiae reformatae, generalis (I. Albertus)

\bibitem[{Houdek {et~al.}(1999)Houdek, {Balmforth}, {Christensen-Dalsgaard}, \&
  {Gough}}]{houdek99}
Houdek, G., {Balmforth}, N.~J., {Christensen-Dalsgaard}, J., \& {Gough}, D.~O.
  1999, \aap, 351, 582

\bibitem[{{Howell} {et~al.}(2012){Howell}, {Rowe}, {Bryson}, {Quinn}, {Marcy},
  {Isaacson}, {Ciardi}, {Chaplin}, {Metcalfe}, {Monteiro}, {Appourchaux},
  {Basu}, {Creevey}, {Gilliland}, {Quirion}, {Stello}, {Kjeldsen},
  {Christensen-Dalsgaard}, {Elsworth}, {Garc{\'{\i}}a}, {Houdek}, {Karoff},
  {Molenda-{\.Z}akowicz}, {Thompson}, {Verner}, {Torres}, {Fressin}, {Crepp},
  {Adams}, {Dupree}, {Sasselov}, {Dressing}, {Borucki}, {Koch}, {Lissauer},
  {Latham}, {Buchhave}, {Gautier}, {Everett}, {Horch}, {Batalha}, {Dunham},
  {Szkody}, {Silva}, {Mighell}, {Holberg}, {Ballot}, {Bedding}, {Bruntt},
  {Campante}, {Handberg}, {Hekker}, {Huber}, {Mathur}, {Mosser}, {R{\'e}gulo},
  {White}, {Christiansen}, {Middour}, {Haas}, {Hall}, {Jenkins}, {McCaulif},
  {Fanelli}, {Kulesa}, {McCarthy}, \& {Henze}}]{Howell2012}
{Howell}, S.~B., {Rowe}, J.~F., {Bryson}, S.~T., {Quinn}, S.~N., {Marcy},
  G.~W., {Isaacson}, H., {Ciardi}, D.~R., {Chaplin}, W.~J., {Metcalfe}, T.~S.,
  {Monteiro}, M.~J.~P.~F.~G., {Appourchaux}, T., {Basu}, S., {Creevey}, O.~L.,
  {Gilliland}, R.~L., {Quirion}, P.-O., {Stello}, D., {Kjeldsen}, H.,
  {Christensen-Dalsgaard}, J., {Elsworth}, Y., {Garc{\'{\i}}a}, R.~A.,
  {Houdek}, G., {Karoff}, C., {Molenda-{\.Z}akowicz}, J., {Thompson}, M.~J.,
  {Verner}, G.~A., {Torres}, G., {Fressin}, F., {Crepp}, J.~R., {Adams}, E.,
  {Dupree}, A., {Sasselov}, D.~D., {Dressing}, C.~D., {Borucki}, W.~J., {Koch},
  D.~G., {Lissauer}, J.~J., {Latham}, D.~W., {Buchhave}, L.~A., {Gautier}, III,
  T.~N., {Everett}, M., {Horch}, E., {Batalha}, N.~M., {Dunham}, E.~W.,
  {Szkody}, P., {Silva}, D.~R., {Mighell}, K., {Holberg}, J., {Ballot}, J.,
  {Bedding}, T.~R., {Bruntt}, H., {Campante}, T.~L., {Handberg}, R., {Hekker},
  S., {Huber}, D., {Mathur}, S., {Mosser}, B., {R{\'e}gulo}, C., {White},
  T.~R., {Christiansen}, J.~L., {Middour}, C.~K., {Haas}, M.~R., {Hall}, J.~R.,
  {Jenkins}, J.~M., {McCaulif}, S., {Fanelli}, M.~N., {Kulesa}, C., {McCarthy},
  D., \& {Henze}, C.~E. 2012, \apj, 746, 123

\bibitem[{{Huber} {et~al.}(2011){Huber}, {Bedding}, {Stello}, {Hekker},
  {Mathur}, {Mosser}, {Verner}, {Bonanno}, {Buzasi}, {Campante}, {Elsworth},
  {Hale}, {Kallinger}, {Silva Aguirre}, {Chaplin}, {De Ridder},
  {Garc{\'{\i}}a}, {Appourchaux}, {Frandsen}, {Houdek}, {Molenda-{\.Z}akowicz},
  {Monteiro}, {Christensen-Dalsgaard}, {Gilliland}, {Kawaler}, {Kjeldsen},
  {Broomhall}, {Corsaro}, {Salabert}, {Sanderfer}, {Seader}, \&
  {Smith}}]{Huber2011}
{Huber}, D., {Bedding}, T.~R., {Stello}, D., {Hekker}, S., {Mathur}, S.,
  {Mosser}, B., {Verner}, G.~A., {Bonanno}, A., {Buzasi}, D.~L., {Campante},
  T.~L., {Elsworth}, Y.~P., {Hale}, S.~J., {Kallinger}, T., {Silva Aguirre},
  V., {Chaplin}, W.~J., {De Ridder}, J., {Garc{\'{\i}}a}, R.~A., {Appourchaux},
  T., {Frandsen}, S., {Houdek}, G., {Molenda-{\.Z}akowicz}, J., {Monteiro},
  M.~J.~P.~F.~G., {Christensen-Dalsgaard}, J., {Gilliland}, R.~L., {Kawaler},
  S.~D., {Kjeldsen}, H., {Broomhall}, A.~M., {Corsaro}, E., {Salabert}, D.,
  {Sanderfer}, D.~T., {Seader}, S.~E., \& {Smith}, J.~C. 2011, \apj, 743, 143

\bibitem[{{Husser} {et~al.}(2013){Husser}, {Wende-von Berg}, {Dreizler},
  {Homeier}, {Reiners}, {Barman}, \& {Hauschildt}}]{Husser2013}
{Husser}, T.-O., {Wende-von Berg}, S., {Dreizler}, S., {Homeier}, D.,
  {Reiners}, A., {Barman}, T., \& {Hauschildt}, P.~H. 2013, \aap, 553, A6

\bibitem[{{Kjeldsen} {et~al.}(1995)}]{Kjeldsen1995}
{Kjeldsen}, H. {et~al.} 1995, A\&A, 293

\bibitem[{{Kraus} \& {Hillenbrand}(2007)}]{Kraus2007}
{Kraus}, A.~L. \& {Hillenbrand}, L.~A. 2007, \aj, 134, 2340

\bibitem[{{Kraus} {et~al.}(2016){Kraus}, {Ireland}, {Huber}, {Mann}, \&
  {Dupuy}}]{Kraus2016}
{Kraus}, A.~L., {Ireland}, M.~J., {Huber}, D., {Mann}, A.~W., \& {Dupuy}, T.~J.
  2016, \aj, 152, 8

\bibitem[{{Lafreni{\`e}re} {et~al.}(2007){Lafreni{\`e}re}, {Marois}, {Doyon},
  {Nadeau}, \& {Artigau}}]{lafreniere07}
{Lafreni{\`e}re}, D., {Marois}, C., {Doyon}, R., {Nadeau}, D., \& {Artigau},
  {\'E}. 2007, \apj, 660, 770

\bibitem[{{Lai}(1997)}]{Lai1997}
{Lai}, D. 1997, \apj, 490, 847

\bibitem[{{Lantz} {et~al.}(2004){Lantz}, {Aldering}, {Antilogus}, {Bonnaud},
  {Capoani}, {Castera}, {Copin}, {Dubet}, {Gangler}, {Henault}, {Lemonnier},
  {Pain}, {Pecontal}, {Pecontal}, \& {Smadja}}]{Lantz2004}
{Lantz}, B., {Aldering}, G., {Antilogus}, P., {Bonnaud}, C., {Capoani}, L.,
  {Castera}, A., {Copin}, Y., {Dubet}, D., {Gangler}, E., {Henault}, F.,
  {Lemonnier}, J.-P., {Pain}, R., {Pecontal}, A., {Pecontal}, E., \& {Smadja},
  G. 2004, in \procspie, Vol. 5249, Optical Design and Engineering, ed.
  L.~{Mazuray}, P.~J. {Rogers}, \& R.~{Wartmann}, 146--155

\bibitem[{{Law} {et~al.}(2014){Law}, {Morton}, {Baranec}, {Riddle},
  {Ravichandran}, {Ziegler}, {Johnson}, {Tendulkar}, {Bui}, {Burse}, {Das},
  {Dekany}, {Kulkarni}, {Punnadi}, \& {Ramaprakash}}]{Law2014}
{Law}, N.~M., {Morton}, T., {Baranec}, C., {Riddle}, R., {Ravichandran}, G.,
  {Ziegler}, C., {Johnson}, J.~A., {Tendulkar}, S.~P., {Bui}, K., {Burse},
  M.~P., {Das}, H.~K., {Dekany}, R.~G., {Kulkarni}, S., {Punnadi}, S., \&
  {Ramaprakash}, A.~N. 2014, \apj, 791, 35

\bibitem[{{Mann} {et~al.}(2015){Mann}, {Feiden}, {Gaidos}, {Boyajian}, \& {von
  Braun}}]{Mann2015}
{Mann}, A.~W., {Feiden}, G.~A., {Gaidos}, E., {Boyajian}, T., \& {von Braun},
  K. 2015, \apj, 804, 64

\bibitem[{{Mann} {et~al.}(2013){Mann}, {Gaidos}, \& {Ansdell}}]{Mann2013}
{Mann}, A.~W., {Gaidos}, E., \& {Ansdell}, M. 2013, \apj, 779, 188

\bibitem[{{Mathur} {et~al.}(2016){Mathur}, {Huber}, {Batalha}, {Ciardi},
  {Bastien}, {Bieryla}, {Buchhave}, {Cochran}, {Endl}, {Esquerdo}, {Furlan},
  {Howard}, {Howell}, {Isaacson}, {Latham}, {MacQueen}, \&
  {Silva}}]{Mathur2016}
{Mathur}, S., {Huber}, D., {Batalha}, N.~M., {Ciardi}, D.~R., {Bastien}, F.~A.,
  {Bieryla}, A., {Buchhave}, L.~A., {Cochran}, W.~D., {Endl}, M., {Esquerdo},
  G.~A., {Furlan}, E., {Howard}, A.~W., {Howell}, S.~B., {Isaacson}, H.,
  {Latham}, D.~W., {MacQueen}, P.~J., \& {Silva}, D.~R. 2016, ArXiv e-prints

\bibitem[{{Metcalfe} {et~al.}(2014){Metcalfe}, {Creevey}, {Do{\u g}an},
  {Mathur}, {Xu}, {Bedding}, {Chaplin}, {Christensen-Dalsgaard}, {Karoff},
  {Trampedach}, {Benomar}, {Brown}, {Buzasi}, {Campante}, \&
  others.}]{Metcalfe2014}
{Metcalfe}, T.~S., {Creevey}, O.~L., {Do{\u g}an}, G., {Mathur}, S., {Xu}, H.,
  {Bedding}, T.~R., {Chaplin}, W.~J., {Christensen-Dalsgaard}, J., {Karoff},
  C., {Trampedach}, R., {Benomar}, O., {Brown}, B.~P., {Buzasi}, D.~L.,
  {Campante}, \& others. 2014, \apjs, 214, 27

\bibitem[{{Mosser} {et~al.}(2012){Mosser}, {Elsworth}, {Hekker}, {Huber},
  {Kallinger}, {Mathur}, {Belkacem}, {Goupil}, {Samadi}, {Barban}, {Bedding},
  {Chaplin}, {Garc{\'{\i}}a}, {Stello}, {De Ridder}, {Middour}, {Morris}, \&
  {Quintana}}]{Mosser2012}
{Mosser}, B., {Elsworth}, Y., {Hekker}, S., {Huber}, D., {Kallinger}, T.,
  {Mathur}, S., {Belkacem}, K., {Goupil}, M.~J., {Samadi}, R., {Barban}, C.,
  {Bedding}, T.~R., {Chaplin}, W.~J., {Garc{\'{\i}}a}, R.~A., {Stello}, D., {De
  Ridder}, J., {Middour}, C.~K., {Morris}, R.~L., \& {Quintana}, E.~V. 2012,
  \aap, 537, A30

\bibitem[{{Polfliet} \& {Smeyers}(1990)}]{Polfliet1990}
{Polfliet}, R. \& {Smeyers}, P. 1990, \aap, 237, 110

\bibitem[{{Riddle} {et~al.}(2014){Riddle}, {Hogstrom}, {Papadopoulos},
  {Baranec}, \& {Law}}]{Riddle2014}
{Riddle}, R.~L., {Hogstrom}, K., {Papadopoulos}, A., {Baranec}, C., \& {Law},
  N.~M. 2014, in \procspie, Vol. 9152, Software and Cyberinfrastructure for
  Astronomy III, 91521E

\bibitem[{{Riddle} {et~al.}(2015){Riddle}, {Tokovinin}, {Mason}, {Hartkopf},
  {Roberts}, {Baranec}, {Law}, {Bui}, {Burse}, {Das}, {Dekany}, {Kulkarni},
  {Punnadi}, {Ramaprakash}, \& {Tendulkar}}]{Riddle2015}
{Riddle}, R.~L., {Tokovinin}, A., {Mason}, B.~D., {Hartkopf}, W.~I., {Roberts},
  Jr., L.~C., {Baranec}, C., {Law}, N.~M., {Bui}, K., {Burse}, M.~P., {Das},
  H.~K., {Dekany}, R.~G., {Kulkarni}, S., {Punnadi}, S., {Ramaprakash}, A.~N.,
  \& {Tendulkar}, S.~P. 2015, \apj, 799, 4

\bibitem[{{Salama} {et~al.}(2016){Salama}, {Baranec}, {Jensen-Clem}, {Riddle},
  {Duev}, {Kulkarni}, \& {Law}}]{Salama2016}
{Salama}, M., {Baranec}, C., {Jensen-Clem}, R., {Riddle}, R., {Duev}, D.,
  {Kulkarni}, S., \& {Law}, N.~M. 2016, in \procspie, Vol. 9909, Society of
  Photo-Optical Instrumentation Engineers (SPIE) Conference Series, 99091A

\bibitem[{{Schwarzenberg-Czerny} {et~al.}(2010){Schwarzenberg-Czerny}, {Weiss},
  {Moffat}, {Zee}, {Rucinski}, {Mochnacki}, {Matthews}, {Breger}, {Kuschnig},
  {Koudelka}, {Orleanski}, {Pamyatnykh}, {Pigulski}, \&
  {Grant}}]{Schwarzenberg-Czerny2010}
{Schwarzenberg-Czerny}, A., {Weiss}, W., {Moffat}, A., {Zee}, R.~E.,
  {Rucinski}, S., {Mochnacki}, S., {Matthews}, J., {Breger}, M., {Kuschnig},
  R., {Koudelka}, O., {Orleanski}, P., {Pamyatnykh}, A., {Pigulski}, A., \&
  {Grant}, C. 2010, in COSPAR Meeting, Vol.~38, 38th COSPAR Scientific
  Assembly, 15

\bibitem[{{Springer} \& {Shaviv}(2013)}]{Springer2013}
{Springer}, O.~M. \& {Shaviv}, N.~J. 2013, \mnras, 434, 1869

\bibitem[{{Stello} {et~al.}(2013){Stello}, {Huber}, {Bedding}, {Benomar},
  {Bildsten}, {Elsworth}, {Gilliland}, {Mosser}, {Paxton}, \&
  {White}}]{Stello2013}
{Stello}, D., {Huber}, D., {Bedding}, T.~R., {Benomar}, O., {Bildsten}, L.,
  {Elsworth}, Y.~P., {Gilliland}, R.~L., {Mosser}, B., {Paxton}, B., \&
  {White}, T.~R. 2013, \apjl, 765, L41

\bibitem[{{Stello} {et~al.}(2011){Stello}, {Huber}, {Kallinger}, {Basu},
  {Mosser}, {Hekker}, {Mathur}, {Garc{\'{\i}}a}, {Bedding}, {Kjeldsen},
  {Gilliland}, {Verner}, {Chaplin}, {Benomar}, {Meibom}, {Grundahl},
  {Elsworth}, {Molenda-{\.Z}akowicz}, {Szab{\'o}}, {Christensen-Dalsgaard},
  {Tenenbaum}, {Twicken}, \& {Uddin}}]{stello11}
{Stello}, D., {Huber}, D., {Kallinger}, T., {Basu}, S., {Mosser}, B., {Hekker},
  S., {Mathur}, S., {Garc{\'{\i}}a}, R.~A., {Bedding}, T.~R., {Kjeldsen}, H.,
  {Gilliland}, R.~L., {Verner}, G.~A., {Chaplin}, W.~J., {Benomar}, O.,
  {Meibom}, S., {Grundahl}, F., {Elsworth}, Y.~P., {Molenda-{\.Z}akowicz}, J.,
  {Szab{\'o}}, R., {Christensen-Dalsgaard}, J., {Tenenbaum}, P., {Twicken},
  J.~D., \& {Uddin}, K. 2011, \apjl, 737, L10

\bibitem[{{Verner} {et~al.}(2011){Verner}, {Chaplin}, {Basu}, {Brown},
  {Hekker}, {Huber}, {Karoff}, {Mathur}, {Metcalfe}, {Mosser}, {Quirion},
  {Appourchaux}, {Bedding}, {Bruntt}, {Campante}, {Elsworth}, {Garc{\'{\i}}a},
  {Handberg}, {R{\'e}gulo}, {Roxburgh}, {Stello}, {Christensen-Dalsgaard},
  {Gilliland}, {Kawaler}, {Kjeldsen}, {Allen}, {Clarke}, \&
  {Girouard}}]{Verner2011}
{Verner}, G.~A., {Chaplin}, W.~J., {Basu}, S., {Brown}, T.~M., {Hekker}, S.,
  {Huber}, D., {Karoff}, C., {Mathur}, S., {Metcalfe}, T.~S., {Mosser}, B.,
  {Quirion}, P.-O., {Appourchaux}, T., {Bedding}, T.~R., {Bruntt}, H.,
  {Campante}, T.~L., {Elsworth}, Y., {Garc{\'{\i}}a}, R.~A., {Handberg}, R.,
  {R{\'e}gulo}, C., {Roxburgh}, I.~W., {Stello}, D., {Christensen-Dalsgaard},
  J., {Gilliland}, R.~L., {Kawaler}, S.~D., {Kjeldsen}, H., {Allen}, C.,
  {Clarke}, B.~D., \& {Girouard}, F.~R. 2011, \apjl, 738, L28

\bibitem[{{Walker} {et~al.}(2003){Walker}, {Matthews}, {Kuschnig}, {Johnson},
  {Rucinski}, {Pazder}, {Burley}, {Walker}, {Skaret}, {Zee}, {Grocott},
  {Carroll}, {Sinclair}, {Sturgeon}, \& {Harron}}]{Walker2003}
{Walker}, G., {Matthews}, J., {Kuschnig}, R., {Johnson}, R., {Rucinski}, S.,
  {Pazder}, J., {Burley}, G., {Walker}, A., {Skaret}, K., {Zee}, R., {Grocott},
  S., {Carroll}, K., {Sinclair}, P., {Sturgeon}, D., \& {Harron}, J. 2003,
  \pasp, 115, 1023

\bibitem[{{Wang} {et~al.}(2014){Wang}, {Fischer}, {Xie}, \&
  {Ciardi}}]{Wang2014}
{Wang}, J., {Fischer}, D.~A., {Xie}, J.-W., \& {Ciardi}, D.~R. 2014, \apj, 791,
  111

\bibitem[{{Ziegler} {et~al.}(2016){Ziegler}, {Law}, {Baranec}, {Morton},
  {Riddle}, {Atkinson}, \& {Nofi}}]{Ziegler2016}
{Ziegler}, C., {Law}, N.~M., {Baranec}, C., {Morton}, T., {Riddle}, R.,
  {Atkinson}, D., \& {Nofi}, L. 2016, ArXiv e-prints

\end{thebibliography}


\clearpage
\begin{sidewaystable}
    \centering
\begin{longtable}{l || c c c c c c c c c}
\caption{Detected Companion Systems \label{CompDet}}\\

\hline\\

\textbf{KIC ID} & \textbf{Separation (\arcsec)} & \textbf{Position} & \textbf{Magnitude} & \textbf{Magnitude} & \textbf{i\textquotesingle{} Detection} & \textbf{Spectral} & \textbf{Amplitude} & \textbf{Amplitude} & \textbf{IR Data}\\
\textbf{} && \textbf{Angle ($^{\circ}$)} & \textbf{Difference i\textquotesingle{}} & \textbf{Difference K\textquotesingle{}} & \textbf{Significance ($\sigma$)} & \textbf{Type\footnotemark[2]} & \textbf{Dilution i\textquotesingle{} (\%)} & \textbf{Dilution K\textquotesingle{} (\%)} &\\

\\


\textbf{5955122} & 3.60 $\pm$ 0.06 & 257.7 $\pm$ 1.6 & 5.15 $\pm$ 0.01 & 4.30 $\pm$ 0.02 & 4.46 & F9IV-V & 0.86 $\pm$ 0.01 & 1.87 $\pm$ 0.05 & Subaru\\
&&&&&&&&&\\
\\

\textbf{7584900} & 2.24 $\pm$ 0.06 & 39.2 $\pm$ 1.7 & 2.40 $\pm$ 0.01 & 2.61 $\pm$ 0.04\footnotemark[1] & 21.9 & K3IV\footnotemark[3] & 9.90 $\pm$ 0.11 & 8.26 $\pm$ 0.37 & Keck\\
&&&&&&&&&\\
\\

\textbf{7747078} & 2.65 $\pm$ 0.06 & 259.8 $\pm$ 1.6 & 2.92 $\pm$ 0.01 & 2.86 $\pm$ 0.02 & 27.1 & F9IV-V & 6.35 $\pm$ 0.10 & 6.71 $\pm$ 0.15 & Subaru\\
&&&&&&&&&\\
\\

\textbf{8394589} & 1.95 $\pm$ 0.06 & 198.0 $\pm$ 1.8 & 3.63 $\pm$ 0.01 & 2.36 $\pm$ 0.02 & 8.86 & F8V & 3.41 $\pm$ 0.05 & 10.2 $\pm$ 0.29 & Subaru \\
&&&&&&&&&\\
\\

\textbf{9139163} & 3.58 $\pm$ 0.06 & 291.7 $\pm$ 1.6 & 6.74 $\pm$ 0.01 & 4.77 $\pm$ 0.02 & 6.99 & F8IV & 0.20 $\pm$ 0.004 & 1.23 $\pm$ 0.03 & Subaru \\
&&&&&&&&&\\
\\

\textbf{9206432} & 3.67 $\pm$ 0.06 & 21.4 $\pm$ 1.6 & 5.92 $\pm$ 0.01 & 5.61 $\pm$ 0.01\footnotemark[1] & 4.81 & F8IV & 0.43 $\pm$ 0.003 & 0.57 $\pm$ 0.01 & Keck\\
&&&&&&&&&\\
\\

\textbf{11026764} & 1.09 $\pm$ 0.06 & 81.0 $\pm$ 2.2 & 1.85 $\pm$ 0.01 & 2.22 $\pm$ 0.02 & 5.05 & G1V & 15.4 $\pm$ 0.22 & 11.5 $\pm$ 0.24 & Subaru\\
&&&&&&&&&\\
\\

\textbf{11554100} & 1.20 $\pm$ 0.06 & 259.3 $\pm$ 2.1 & 4.04 $\pm$ 0.05 & 4.05 $\pm$ 0.02\footnotemark[1] & \textless3 & K3IV\footnotemark[1] & 2.46 $\pm$ 0.15 & 2.34 $\pm$ 0.07 & Keck\\
\textbf{} &&&&&&&&&\\
\\
\hline
\end{longtable}

\protect\footnotetext[1]{Filter: Br$\gamma$.}


\protect\footnotetext[2]{Values taken from KASOC: http://kasoc.phys.au.dk.} 

\protect\footnotetext[3]{Values not found on KASOC. Individual magnitude values for J, H, K\textquotesingle{} and i\textquotesingle{} taken from CFOP: https://exofop.ipac.caltech.edu/cfop.php and used in conjunction with the code from \cite{Atkinson2016} to determine a composite type.}

\protect\footnotetext[4]{Composite value of system not available on CFOP. J, H and K\textquotesingle{} values from CFOP used with \cite{Atkinson2016} models to determine i\textquotesingle{}.}

\end{sidewaystable}

\clearpage
\begin{sidewaystable}
    \centering
\begin{longtable}{l || c c c c c c c c c c }
\caption{Individual Star Information for Companion Systems \label{CompDet2}}\\

\hline\\

\textbf{KIC ID} & \textbf{Primary} & \textbf{Secondary} & \textbf{Spectral} & \textbf{Spectral} &  \textbf{Radius} & \textbf{Distance} & \textbf{Error\footnotemark[4]} & \textbf{Distance} & \textbf{Error} & \textbf{Overlap} \\
\textbf{} & \textbf{Magnitude} & \textbf{Magnitude} & \textbf{Type} & \textbf{Type} & \textbf{Secondary} & \textbf{Primary\footnotemark[3]} && \textbf{Secondary} & & \textbf{Within} \\
\textbf{} & \textbf{i\textquotesingle{} and K\textquotesingle{}} & \textbf{i\textquotesingle{} and K\textquotesingle{}} & \textbf{Primary} & \textbf{Secondary} & \textbf{(R$\odot$)} & \textbf{(pc)} & & \textbf{(pc)}  & & \textbf{Error} \\

\\


\textbf{5955122} & 9.13 $\pm$ 0.03 & 14.3 $\pm$ 0.03 & F7V & M4V\footnotemark[1] & 0.88 $\pm$ 0.01 & 169.0 & + 8.4 & 80\footnotemark[1] & + 20 & NO \\\
& 7.94 $\pm$ 0.05 & 12.2 $\pm$ 0.05 & & & && - 7.6 & & - 20 & 4.13$\sigma$ \\
\\

\textbf{7584900} & 11.1 $\pm$ 0.02 & 13.5 $\pm$ 0.02 & K3IV & K2V\footnotemark[2] & 0.93 $\pm$ 0.02 & 499.3 & + 32.5 & 339 & + 32 & $\dddot{}$\\
& 9.19 $\pm$ 0.08 & 11.8 $\pm$ 0.08 & & & &&  - 51.1 & & - 24 & 2.24$\sigma$ \\
\\

\textbf{7747078} & 9.25 $\pm$ 0.03 & 12.2 $\pm$ 0.03 & F7V & G3 & 1.11 $\pm$ 0.08 & 170.3 & + 7.4 & 340 & + 53 & NO \\
& 8.05 $\pm$ 0.04 & 10.9 $\pm$ 0.04 & & & &&  - 8.2 & & - 43 & 4.01$\sigma$ \\
\\

\textbf{8394589} & 9.43 $\pm$ 0.03 & 13.1 $\pm$ 0.03 & F5V & K8V\footnotemark[2] & 0.79 $\pm$ 0.01 & 116.9 & + 5.2 & 147 & + 7 & NO \\\
& 8.34 $\pm$ 0.05 & 10.7 $\pm$ 0.05 & & & &&  - 5.2 & & - 7 & 3.69$\sigma$ \\
\\

\textbf{9139163} & 8.21 $\pm$ 0.04 & 15.0 $\pm$ 0.04 & F3IV-V & M3 & 0.42 $\pm$ 0.04 & 97.4 & + 8.5 & 150 & + 20 & $\dddot{}$ \\
& 7.24 $\pm$ 0.04 & 12.0 $\pm$ 0.04 & & & &&  - 8.5 & & - 19 & 2.53$\sigma$ \\
\\

\textbf{9206432} & 8.98 $\pm$ 0.01 & 14.9 $\pm$ 0.01 & F2IV-V & F9 & 1.17 $\pm$ 0.04 & 137.2 & + 4.1 & 1300 & + 120 & NO \\
& 8.07 $\pm$ 0.03 & 13.7 $\pm$ 0.03 & & & &&  - 5.5 & & - 170 & 6.84$\sigma$ \\
\\

\textbf{11026764} & 9.29 $\pm$ 0.03 & 11.2 $\pm$ 0.03 & G4V & F3V & 1.33 $\pm$ 0.03 & 172.1 & + 16.2 & 393 & + 31 & NO \\
& 8.00 $\pm$ 0.04 & 10.2 $\pm$ 0.04 & & & &&  - 16.2 & & - 29 & 7.07$\sigma$ \\
\\

\textbf{11554100} & 11.1 $\pm$ 0.11 & 15.1 $\pm$ 0.11 & K3IV & K3 & 0.91 $\pm$ 0.02 & 442.5 & + 75.8 & 632 & + 50 & $\dddot{}$ \\
& 9.25 $\pm$ 0.05 & 13.3 $\pm$ 0.05 & & & &&  - 21.9 & & - 33 & 2.36$\sigma$ \\
\\

\hline
\end{longtable}


\protect\footnotetext[1]{Value from the spectroscopic data.}

\protect\footnotetext[2]{Spectroscopic values corroborate with the photometric values here.}

\protect\footnotetext[3]{Corrected for the effects of amplitude dilution.}

\protect\footnotetext[4]{Errors include propagation of amplitude dilution errors.}
\end{sidewaystable}

\clearpage
\clearpage
\begin{longtable}[c]{ l  c  c  c  c  c }
 
\caption{Full Robo-AO Observation List \label{long}}\\[0.5ex]

\hline
\\
\textbf{KIC ID} & \textbf{m\textsubscript{i\textquotesingle{}}(mags)} & \textbf{Obs Date} & \textbf{Companion?} & \textbf{KOI}\\[0.5ex]

\hline
\endfirsthead

\multicolumn{6}{c}{\textbf{Table \ref{long} Continued}}\\[0.5ex]
\hline
\\
\textbf{KIC ID} & \textbf{m\textsubscript{i\textquotesingle{}}(mags)} & \textbf{Obs Date} & \textbf{Companion?} & \textbf{KOI}\\[0.5ex]

\hline\\
\endhead

\endlastfoot
\startdata
\\
1435467&8.7&2014 Jul 13&& \\ [0.5ex]
2837475&8.4&2014 Jul 13&& \\ [0.5ex]
2852862&10.6&2014 Jul 13&&\\[0.5ex]
3424541&9.6&2014 Jul 11&&\\[0.5ex]
3427720&9.0&2014 Jul 11&&\\[0.5ex]
3632418&...&2014 Jul 13&A$+$&K00975.01\\[0.5ex]
3656476&9.3&2014 Jul 13&&\\[0.5ex]
3735871&9.6&2014 Jul 13&&\\[0.5ex]
4351319&9.6&2014 Jul 13&&\\[0.5ex]
4914923&...&2014 Jul 13&&\\[0.5ex]
5184732&8.0&2014 Jul 13&&\\[0.5ex]
5596656&9.2&2014 Jul 18&&\\[0.5ex]
5607242&10.5&2014 Jul 13&&\\[0.5ex]
5689820&11.0&2014 Jul 13&&\\[0.5ex]
5723165&10.3&2014 Aug 28&&\\[0.5ex]
5955122&9.1&2014 Jul 13&YES&\\[0.5ex]
6064910&11.3&2014 Jul 12&&\\[0.5ex]
6106415&...&2014 Sep 1&&\\[0.5ex]
6116048&8.3&2014 Jul 13&&\\[0.5ex]
6225718&...&2014 Aug 20&&\\[0.5ex]
6442183&...&2014 Jul 13&&\\[0.5ex]
6531928&10.4&2014 Jul 13&&\\[0.5ex]
6603624&8.9&2014 Jul 13&&\\[0.5ex]
6679371&8.7&2014 Jul 13&&\\[0.5ex]
6693861&11.3&2014 Sep 2&&&\\[0.5ex]
6766513&11.2&2014 Jul 13&&\\[0.5ex]
6933899&9.4&2014 Sep 1&&\\[0.5ex]
7103006&8.8&2014 Jul 13&&\\[0.5ex]
7107778&11.1&2014 Jul 13&&\\[0.5ex]
7206837&9.7&2014 Jul 13&&\\[0.5ex]
7296438&9.9&2014 Jul 12&CFOP&K00364.01\\[0.5ex]
7341231&...&2014 Jul 18&&\\[0.5ex]
7584900&11.0&2014 Jun 19&YES&\\[0.5ex]
7680114&9.9&2014 Jul 13&&\\[0.5ex]
7747078&...&2014 Jul 13&YES&\\[0.5ex]
7771282&10.6&2014 Jul 13&&\\[0.5ex]
7799349&9.2&2014 Sep 2&&\\[0.5ex]
7800289&9.4&2014 Jun 19&&\\[0.5ex]
7871531&9.0&2014 Jun 19&&\\[0.5ex]
7970740&7.6&2014 Aug 28&&\\[0.5ex]
7976303&8.9&2014 Jul 13&&\\[0.5ex]
8006161&7.1&2014 Jun 17&&\\[0.5ex]
8018599&...&2014 Jul 13&&\\[0.5ex]
8179536&9.4&2014 Jul 13&&\\[0.5ex]
8179973&10.0&2014 Jul 13&CFOP&K01019.01\\[0.5ex]
8228742&9.2&2014 Nov 7&&\\[0.5ex]
8394589&9.4&2014 Aug 21&YES&\\[0.5ex]
8424992&10.1&2014 Jul 13&&\\[0.5ex]
8524425&9.5&2014 Jul 11&&\\[0.5ex]
8561221&9.6&2014 Jul 13&&\\[0.5ex]
8694723&8.7&2014 Jul 13&&\\[0.5ex]
8702606&9.1&2014 Jul 13&&\\[0.5ex]
8751420&...&2014 Jul 13&&\\[0.5ex]
8760414&...&2014 Jul 13&&\\[0.5ex]
9025370&8.7&2014 Jul 13&&\\[0.5ex]
9073950&11.1&2014 Jul 13&&\\[0.5ex]
9098294&9.6&2014 Jul 13&&\\[0.5ex]
9139151&9.0&2014 Jul 13&&\\[0.5ex]
9139163&8.2&2014 Jul 13&YES&\\[0.5ex]
9206432&9.0&2014 Jul 13&YES&\\[0.5ex]
9353712&...&2014 Jul 13&&\\[0.5ex]
9410862&10.6&2014 Jul 13&&\\[0.5ex]
9414417&9.5&2014 Jul 13&&K00974.01\\[0.5ex]
9574283&10.5&2014 Jul 13&&\\[0.5ex]
9812850&9.4&2014 Jul 13&&\\[0.5ex]
9955598&9.2&2014 Jul 13&&K01925.01\\[0.5ex]
10018963&8.6&2014 Jul 13&&\\[0.5ex]
10068307&8.1&2014 Jul 13&&\\[0.5ex]
10079226&9.9&2014 Jul 13&&\\[0.5ex]
10130853&10.6&2014 Jul 13&&\\[0.5ex]
10147635&10.5&2014 Jul 13&&\\[0.5ex]
10162436&8.5&2014 Jul 12&&\\[0.5ex]
10272858&11.1&2014 Sep 2&&K05782.01\\[0.5ex]
10454113&8.5&2014 Jul 13&&\\[0.5ex]
10516096&9.3&2014 Jul 17&&\\[0.5ex]
10593351&10.6&2014 Jul 13&&\\[0.5ex]
10644253&9.0&2014 Jun 15&&\\[0.5ex]
10709834&9.7&2014 Jun 17&&\\[0.5ex]
10963065&8.7&2014 Sep 3&&K01612.01\\[0.5ex]
10972873&10.5&2014 Jul 13&&\\[0.5ex]
11026764&9.1&2014 Jul 13&YES&\\[0.5ex]
11081729&9.0&2014 Jul 13&&\\[0.5ex]
11137075&10.7&2014 Jul 13&&\\[0.5ex]
11193681&10.5&2014 Jul 13&&\\[0.5ex]
11244118&9.5&2014 Jul 13&&\\[0.5ex]
11253226&8.4&2014 Jul 13&&\\[0.5ex]
11414712&...&2014 Jul 13&&\\[0.5ex]
11554100&8.4&2014 Jul 13&YES&\\[0.5ex]
11717120&11.1&2014 Jul 13&&\\[0.5ex]
11771760&9.0&2014 Jul 13&&\\[0.5ex]
11968749&11.2&2014 Jul 13&&\\[0.5ex]
12009504&10.1&2014 Jul 13&&\\[0.5ex]
12069127&9.2&2014 Jul 13&&\\[0.5ex]
12069424&10.5&2014 Jul 13&&\\[0.5ex]
12069449&...&2014 Jul 13&&\\[0.5ex]
12258514&...&2014 Jul 13&&\\[0.5ex]
12307366&11.2&2014 Jul 13&&\\[0.5ex]
12317678&8.6&2014 Jul 13&&\\[0.5ex]
12508433&9.3&2014 Jul 13&&\\[0.5ex]
\end{longtable}
Notes. --- References for previous detections are denoted using the following codes: (A$+$) for \citep{Adams2012, Ginski2016, Howell2012, Wang2014, Kraus2016}, and (CFOP) for high angular resolution images available on \textit{Kepler} Community FollowUp Observing Program.\\

\clearpage
\begin{longtable}[c]{c  c  c  c  c}
 
\caption{Full IR Observations\label{irobs}}\\[0.5ex]

\hline
\\
\textbf{KIC ID} & \textbf{Obs Date} & \textbf{Instrument} & \textbf{Total Exposure} & \textbf{Companion?}\\[0.5ex]
&&&\textbf{Time (secs)}& \\


\hline\\
\endhead


2852862 & 2016 Apr 16 & NIRC2 & 45 & $\dddot{}$\\[0.5ex]
3735871 & 2016 Apr 16 & NIRC2 & 90 & $\dddot{}$\\[0.5ex]
5184732 & 2016 Apr 16 & NIRC2 & 45 & $\dddot{}$\\[0.5ex]
5955122 & 2016 Jun 17 & IRCS & 1680 & YES\\[0.5ex]
7747078 & 2016 Jun 17 & IRCS & 4.12 & YES\\[0.5ex]
8394589 & 2016 Jun 17 & IRCS & 944 & YES\\[0.5ex]
9139163 & 2016 Jun 17 & IRCS & 472 & YES\\[0.5ex]
11026764 & 2016 Jun 17 & IRCS & 2.45 & YES\\[0.5ex]
11554100 & 2016 Apr 16 & NIRC2 & 45 & YES\\[0.5ex]
7584900 & 2016 Sep 13 & NIRC2 & 180 & YES\\[0.5ex]
9206432   & 2016 Jun 17 & IRCS & 1 & $\dddot{}$ \\[0.5ex]
& 2016 Sep 12 & NIRC2 & 120 & YES\\[0.5ex]

\end{longtable}

\end{document}